\documentclass[lettersize,journal]{IEEEtran}
\usepackage{amsmath,amsfonts}
\usepackage{algorithmic}
\usepackage{array}
\usepackage{textcomp}
\usepackage{stfloats}
\usepackage{url}
\usepackage{verbatim}
\usepackage{graphicx}
\usepackage{cite}
\usepackage[T1]{fontenc}
\usepackage{tabularx}
\usepackage{graphicx}
\usepackage[cmintegrals]{newtxmath}
\usepackage{algorithmic}
\usepackage{cite}
\usepackage{setspace} 
\usepackage{subcaption}
\usepackage{xcolor}
\usepackage[para]{footmisc}
\usepackage{tabularx,array}
\usepackage{multirow}
\usepackage{booktabs}
\usepackage[linesnumbered,ruled,vlined]{algorithm2e}
\SetKwInput{KwInput}{Input}              
\SetKwInput{KwOutput}{Output}  
\SetKwInput{KwFunction}{Function}  
\SetKwInput{KwReturn}{Return}  
\usepackage{textcomp}
\usepackage{linguex}
\usepackage{metalogo}
\usepackage{url}
\usepackage{csquotes} 
\usepackage{arydshln} 
\usepackage{color, soul}
\usepackage{makecell}
\usepackage{etoolbox}
\usepackage{threeparttable}
\usepackage{pifont}
\usepackage{float}
\usepackage[normalem]{ulem}
\usepackage{mathtools}
\DeclarePairedDelimiter{\ceil}{\lceil}{\rceil}
\usepackage{comment}
\usepackage{balance}

\begin{document}

\title{On Attack-Resilient Service Placement and Availability in Edge-enabled IoV Networks} 

\author{Anum~Talpur,~\IEEEmembership{Member,~IEEE,}
	and~Mohan~Gurusamy,~\IEEEmembership{Senior~Member,~IEEE}% 
	
\thanks{A. Talpur and M. Gurusamy are with the Department of Electrical and Computer Engineering, National University of Singapore, Singapore (email: anum.talpur@u.nus.edu; gmohan@nus.edu.sg).}}% 
        
\maketitle

\begin{abstract}
Achieving network resilience in terms of attack tolerance and service availability is critically important for Internet of Vehicles (IoV) networks where vehicles require assistance in sensitive and safety-critical applications like driving. It is particularly challenging in time-varying conditions of IoV traffic. In this paper, we study an attack-resilient optimal service placement problem to ensure disruption-free service availability to the users in edge-enabled IoV network. Our work aims to improve the user experience while minimizing the delay and simultaneously considering efficient utilization of limited edge resources. First, an optimal service placement is performed while considering traffic dynamicity and meeting the service requirements with the use of a deep reinforcement learning (DRL) framework. Next, an optimal secondary mapping and service recovery placements are performed to account for the attacks/failures at the edge. The use of DRL framework helps to adapt to dynamically varying IoV traffic and service demands. In this work, we develop three ILP models and use them in the DRL based framework to provide attack-resilient service placement and ensure service availability with efficient network performance. Extensive numerical experiments are performed to demonstrate the effectiveness of the proposed approach.
\end{abstract}


\section{Introduction}
Whether it's freeing humans from tedious driving tasks or minimizing the risk of catastrophes by eliminating the possibilities of human errors, the Internet of Vehicles (IoV) integrates a series of network services to help vehicles conduct necessary tasks related to driving and vehicle automation. IoV being a mission-critical application is categorized under 5G ultra-reliable and low latency communications (URLLC) technology \cite{urllc}. This makes low latency and high reliability a key requirement for IoV services. The vehicles need to be connected in real-time with each other and the infrastructure to get assistance in different computation tasks. The use of cloud computing poses a significant challenge as it may not be suitable to process real-time traffic conditions of strict latency and reliability requirements. Edge network (EN) is a potential technology that can assure faster service availability by bringing computing resources closer to the end-users. Therefore, the ENs are anticipated to promise great benefits in building an intelligent Internet of Vehicles (IoV) network \cite{inntroEdge,anumsurvey}. \par 
EN is a network architecture that brings computing resources to the edge of the cellular network with one or more edge nodes or servers \cite{inntroEdge,edgeattacks}. The high versatility of locations and closer positions of edge nodes eliminates long-distance communications. The data that need to be sent between the user and infrastructure can be transmitted faster with low latency. EN also offers the benefit of distributed local processing, caching, and improved security. Despite these advantages, EN is in its early stage of development where several problems need to be addressed. One of the common and important problems is security against different attack types. In this work, we focus on optimal and attack-resilient service placement at EN for disruption-free or minimal-disruption service availability to the vehicles for sensitive applications like driving. \par
Resilience is the capability to maintain a good quality of service in the system in the event of attacks or failures. The problem of providing resilience in ENs is new and not explored much for the application of service placement in an IoV network. Using backup resource (BR) reservation methods are common and widely used in the recent literature to handle service failures \cite{backup2,backup3,backup4,backup5,backup1,backup6}. However, the BR reservation based solutions are resource expensive and result in wastage where resources remain idle until failure occurs. In the application of service placement, the over-provisioning of resources due the backup service instances limits the choice of optimal placement for primary instances and results in longer delays observed by vehicles. Thus, an important research question is how to provide services resilient to attacks on an edge node (server) with minimal number of active edge servers and resources while satisfying service requirements and low or no disruption in the event of a failure. \par 
In this paper, we consider an IoV network where different types of services are offered to the vehicles to help them coordinate in remote driving, road safety, and many other applications. We start with the service placement to find the optimal placement of services at the edge servers while considering the vehicle's mobility and dynamics in the requests. More than one service instances are placed at different edge servers for the purpose of meeting service requirements and also resilience. During the normal operation, a vehicle receives its service from a server based on primary vehicle-to-edge (V2E) mapping and upon the service failure, it needs to get service from an attack-free server chosen a-priori as secondary V2E mapping. While this pro-active secondary mapping ensures low or no service disruption, it might overload the attack-free working servers. Therefore, as the last stage, our design framework chooses a recovery (reactive) solution to find new server(s) for the failed service instances.\par 
Such an attack-resilient service placement framework of multiple objectives needs to satisfy multiple design criteria such as low delay, low service disruption, and high resource efficiency considering the dynamic nature of service demands. Our proposed approach aims to achieve high level of system performance from the perspective of users as well as service providers for service placement with resilience against service failure at an edge server. We develop three different ILP-based formulations to solve the attack-resilient service placement problem and make the following contributions in this paper. \par
\begin{itemize}
	\item First, we perform optimal placement of multiple instances of each service type on edge nodes while guaranteeing minimal service delay and minimal edge resource usage. The effects of dynamic traffic volumes on service placement (SP) are addressed using the deep reinforcement learning (DRL) framework with an ILP model of our earlier work in \cite{DRLD-SP}. This presents the primary V2E mapping for vehicles to avail different services.
	\item Second, we develop an ILP model, to determine proactive secondary V2E mapping (PSVM) to provide low service delay. We note that we do not reserve any resources but only select a secondary server to be used in the event of an attack. Thus, we do not need additional backup resources during the normal operation. Upon an attack, an affected vehicle will switch to the secondary server with no or little service disruption.
	\item Third, we develop an ILP model to address the recovery problem to perform service recovery placement (SRP) for the affected service instances hosted at the edge node under attack. The failed instances are re-instantiated at the unaffected edge servers. The affected vehicles will then switch to these new service instances. This is because, while the PSVM enables an affected vehicle to receive services with minimal disruption, the servers may be congested as no additional resources were reserved during the normal operation.  
	\item Finally, our performance evaluation results present useful insights on service delay, edge resource usage, number of active servers, and run time by our proposed framework against the baseline method based on backup resource (BR) reservation.
\end{itemize}
The rest of the paper is organized as follows. Section \ref{Sec:RelatedWork} presents the related works. Section \ref{Sec:ModelandProblem} provides the details on the system model, service model, and problem description. In Section \ref{Sec:ProblemFormulation}, we develop formulations for three ILP models to solve the problem of resilient service placement. In Section \ref{Sec:ResilientPlacementFramework}, we present the architecture and algorithms for the proposed attack-resilient service placement framework. Section \ref{Sec:PerformanceStudy} performs the performance study, and finally, Section \ref{Sec:Conclusion} makes concluding remarks.

\section{Related Work}
\label{Sec:RelatedWork}
The diversified service requirements of vehicular applications ranging from time-sensitive remote driving to resource-intensive autonomous driving video processing, foster the inclusion of EN with IoV architecture. The EN which processes data closer and the first step between the user and core of the network is also a target for attacks. In the literature, there are several survey works presented recently on the security and privacy issues of network edge computing \cite{edgeattacks,edgeattacks2,edgeattacks3}. The ENs are comprised of various distributed entities like wireless networks, storage networks, visualization system programs, and so on. The above survey works discuss the possible attacks on different components of the EN. Different kinds of attacks and their impacts are discussed where the attacks like denial-of-service (DoS) attacks, jamming attacks, and forgery attacks may fail edge nodes or perform system damage. There also exist several works which use edge capabilities to provide defense against attacks over IoT devices or cloud networks \cite{useEdge1,useEdge2}. However, the failure of edge node due to an attack has not much been studied in the literature. \par
He et al. in a recent work in \cite{gametheoryDOSdetection} use a game-theoretic approach to formulate a decentralized algorithm and find the Nash equilibrium as a solution to the edge node attack problem. The idea is to find allocation decisions in a way that mitigation cost is minimum and the edge server does not exceed its processing capacity. The mitigation cost here is in terms of the number of hops. This work does not deal with the mobile traffic of an IoV network and does not optimize the performance parameters which are crucial for IoVs. The data used in this work is generated following Poisson distribution which is not realistic. Singh et al. in \cite{GaussianDoSedgeAttack} detects DoS attacks on EN using a Bayesian classifier to decide whether the new packet is from a legitimate user or not. This work focuses on detection only and doesn't provide a solution for prevention or attack recovery. Another work in \cite{userMigration} deals with the problem of edge server failure as an integer programming problem. This work proposes an optimal strategy for migration of end-users to nearby edge servers and to maintain service delivery. This work is only delay focused and doesn't consider to recover the failed system resources. The dynamics in traffic volumes are also not considered in this work. \par
Different from the above works, we study the problem of attack-resilient service placement in edge-enabled IoV networks. The problem of resilient service placement in IoV is new and not widely explored. Some recent works study the problem of resilient service placement for different applications. Using redundancy and advance reservation of backup resources is the most common and well-known resiliency method used to handle node failures \cite{backup2,backup3,backup4,backup5,backup1}. The redundant backups have their drawbacks including resource wastage, higher cost, and higher delays. The authors in different recent works \cite{backup4,backup6} propose the techniques to cut down or reduce the backup edge resources. Despite this, the drawbacks of redundancy or backup are inevitable. \par 
Kang et al. \cite{Kang} propose a mixed integer programming model to perform resilient resource allocation in service function chains by selecting suitable replicas from the pools of replicas and deciding the deployment locations of these replicas. This work considers the problem of node failures where virtual network functions (VNFs) running on the node also fail but the node has a copy of data and status information through which replicas would be generated and replaced. Their objective is to minimize the recovery time. This work is VNF-focused and doesn't study the edge network scenarios. Another work on resilient VNF placement is presented in \cite{Purnima}. This work considers the use of cloud resources along with edge while proposing their solution. However, the use of cloud may not be an optimal choice for delay-sensitive IoV applications. \par
Cheng et al. \cite{Cheng} consider edge nodes only and present a resilient service placement under uncertainty. This work proposes an ILP model with the objective of cost-optimal placement. The cost is in dollars which the user pays based on per-hour resource usage. The authors of this work study one service per edge node scenario and do not consider the mobile traffic which makes it unsuitable for IoV applications. The performance evaluation is performed on simulated data where link delays are also generated randomly. Yuben et al. propose a resilient service provisioning for edge computing in \cite{Yuben} to maximize the expected overall utility. This work studies the problem of uncertain failures but doesn't consider the mobile traffic. The traffic is static with a fixed number of users considered from a fixed wireless router as an input for performance evaluation.  \par 
There are few works on resilient controller placement presented in the literature \cite{CP1,CP2,vignesh}. But the resilience techniques proposed in these works are not suitable for ENs where resources are limited and also not applicable over mobile traffic of IoV networks. The objective of our work is to propose a resilient edge service placement solution for dynamic and time-varying IoV networks. Note that the dynamicity of the IoV network includes not only the traffic demand, but also the vehicles' geographic locations. Such a dynamicity has an impact on the real-time performance of edge service placement decisions and service availability. 

\section{System Model and Problem Description}
\label{Sec:ModelandProblem}
In this section, we describe the system model, service model, and research problem as the problem description studied in our work.

\subsection{System Model}
\begin{figure}[htbp]
	\centering
	\includegraphics[width=2.1in, height=1.25in]{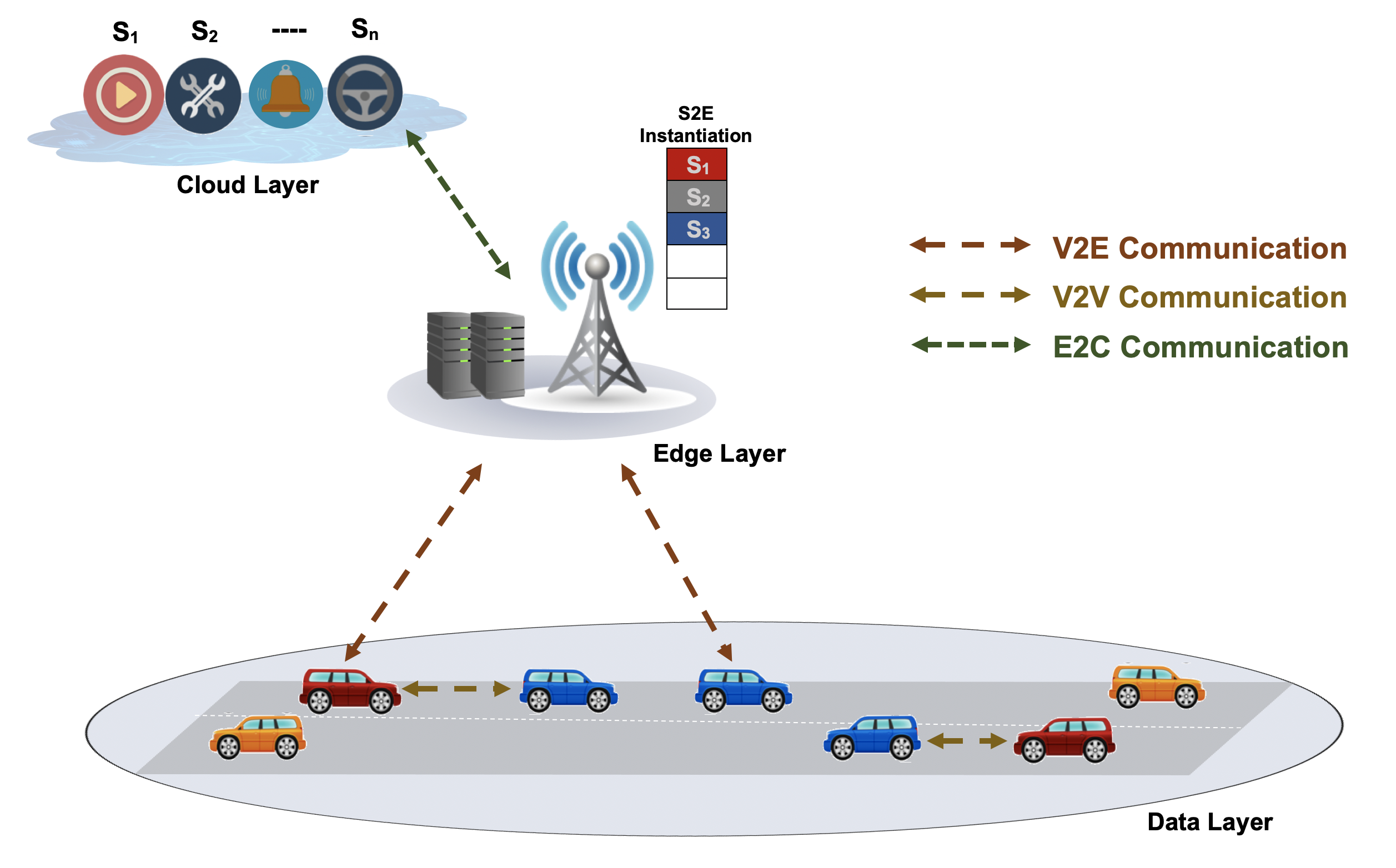}
	\caption{The architecture of edge-enabled IoV network}
	\label{fig:Architecture}
\end{figure}
The architecture of an edge-enabled IoV network is presented in Fig. \ref{fig:Architecture} which is comprised of three layers, namely, data, edge, and cloud. At the \textit{data layer}, we consider a city road environment with taxis (vehicles) travelling in multiple lanes in different directions in the city San Francisco \cite{sanfrancisco}. The vehicles are mobile and equipped with necessary sensors which enable them to provide relevant information.  Vehicles can connect with the edge layer as "V2E mapping" to avail different types of services to get assistance in different tasks related to driving. The service model is further explained in detail in Section \ref{Sec:ServiceModel}. \par
For the \textit{edge layer}, we consider a 5G edge network comprised of edge nodes using evolved NodeB (eNB) stations and forming a multi-cell coverage area for mobile vehicles. Each edge node is equipped with edge servers having limited storage and computation capacity to run different applications and services. For the simplicity purpose, we consider an abstract measure of resources for storage, computation, and memory, for each edge node. We assume each edge node $e\, \epsilon\, E$ has $C_e$ resource units in total. The attacks are assumed to take place at the edge layer. The attack scenario is discussed in Section \ref{Sec:ProblemDefinition}. Additionally, the edge layer connects to the large capacity \textit{cloud layer} via a backbone network to download different service types and perform service-to-edge (S2E) instantiation. We assume adequate links between V2E, E2E (edge-to-edge) nodes/servers, and E2C (edge-to-cloud) are available to enable communication among them.

\subsection{Service Model}
\label{Sec:ServiceModel}
The traffic monitoring and management systems in IoV networks require vehicles to request different types of services to carry out driving tasks and get assistance in emergencies. In this paper, we assume vehicles generate service requests with a certain rate for a type of service. A service is a facility like CAM (cooperative awareness message) service, diagnostic service, environmental notification service, media downloading/sharing service, remote driving service, etc. Each service has a stringent delay requirement $D_s$ and consumes computation and storage resources. For the simplicity purpose, we consider an abstract measure of resources and we assume service type $s\, \epsilon\, S$ requires $R_s$ resource units to instantiate. The service to be deployed on edge nodes comprises multiple service types where each type has multiple service instances (SIs) to meet the requirements and effective provisioning of a service.  \par
Let $S$ be the set of service types and $I_s$ be the number of instances required for service type $s$ to enhance the user experience for that service. A vehicle $v\, \epsilon\, V$ requires a service $s\, \epsilon\, S$ which is to be hosted at an edge node. The number of vehicles requesting service $s$ is uniformly distributed and the arrival rate on edge at time interval $t$ is denoted as $\lambda_s(t)$. Each instance is assumed to have the same processing capacity i.e. the number of vehicles one SI can handle (or provide parallel connections) at a time without queuing delay is denoted by $\mathbb{C}$. Further, a service request is specified as a 4-tuple structure $<v\,,\, l\,,\, t\,,\, s>$. Here, $t$ is the time at which the request is generated and $l$ is the location of the vehicle $v$ requesting for service type $s$. In response to the request, the location of the best edge node/server will be communicated to the vehicles to access the requested services.

\subsection{Problem Description}
\label{Sec:ProblemDefinition}
In this paper, we consider an attack problem that causes an edge node to fail where its computing and communicating capabilities fail. The attack failure scenario causes all SIs running on that node to fail. This failure is either due to a jamming attack, outage attack, denial of service attack, or any other type of attack that causes edge node capabilities to fail completely \cite{edgeattacks}. To endure service provisioning without being permanently failed during an attack, the service placement should fulfil the attack-resilient conditions in addition to other performance metrics. Using redundancy or BR reservation method is a well-known method in the recent literature to handle service failures, as discussed in Section \ref{Sec:RelatedWork}. Maintaining backup has its drawbacks including resource wastage and high service delays which are discussed and compared in detail in the performance study section. In addition, in a vehicular application where traffic is time-varying, the service placement operation has a great impact on the performance of the service requests. Note that the time-varying property of IoV networks is not only related to the dynamic traffic demands, but also the vehicles' geographic locations. Considering these limitations, we propose an attack-resilient service placement framework using ILP formulations with DRL to meet the resilience and address the delay performance and resource limitations of edge-enabled time-varying IoV network.

\section{Problem Formulation}
\label{Sec:ProblemFormulation}
In this section, we formulate the resilient service placement problem using binary ILP. This problem has three parts and accordingly three problems which are explained below. \par 
The first part deals with the optimal service placement problem. The objective is to decide on optimal edge locations to instantiate service instances of different requested service types, subject to resilience, delay and resource constraints. The service requests are generated from vehicles in the specific format described in the previous section, requesting different service types to get assistance in driving tasks. Upon processing a service request, the the details of the chosen edge server hosting that service instance is communicated to the vehicles. The optimal service placement is stated as follows,\par 
\textbf{\textit{Problem 1 (Service Placement (SP)):}} Given a set of service types $s$ and $I_s$ number of service instances for each service type with their resource and delay requirements, the problem is to find the optimal placement $x_e^s$ of services at the edge servers while considering the vehicle's mobility and dynamics in the requests for different types of services. \par
In the second ILP model, we consider the problem of finding two optimal secondary candidate servers (i.e $\kappa_1$ and $\kappa_2$) for V2E mapping following the SP locations decided in problem 1. The discovery of optimal secondary mappings is subject to the lowest delay experienced by vehicles while accessing that service. The idea is to use the secondary mappings when an attack takes place on the node where vehicles are primarily mapped and ensure the continuous service availability for the vehicles. This part of our design aims to minimize the impact of the attack until the system recovers. The decision on secondary candidates in a proactive manner is indeed important consideration to avoid the service disruption that may impact the service availability specially when the service type requested is related to driving and sensitive to delay. We note that while we select the secondary candidates a-priori, we do not reserve any resource. A vehicle will be switched to the first candidate if its service instance (during normal operation) was not placed at the first candidate; otherwise it will be switched to the second candidate. The proactive selection ensures immediate switching of a vehicle to a new server upon service failure. Since the resources are not reserved, the service load on the secondary candidate servers could increase and hence our objective is to minimize the delay. Note that the priority of candidate $\kappa_1$ is greater than $\kappa_2$ considering the delay from $\kappa_1$ is less than $\kappa_2$. Our second problem of optimal proactive secondary V2E mapping  is stated as follows, \par
\textbf{\textit{Problem 2 (Proactive Secondary V2E Mapping (PSVM)):}} Given a set of optimal service placement $x_e^s$ to deploy $I_s$ instances of each service type $s$ with their resource and delay requirements, the problem is to find the two best secondary mappings, i.e. $y_{e,\kappa_1}^s$ and $y_{e,\kappa_2 }^s$, minimizing the delay, and ensure continuous service availability for the set of vehicles $\mathcal{V'}$  impacted by the attack. \par 
In the third and last optimization problem, we perform the recovery/re-instantiation of attacked SIs. We carry out the re-instantiation of attacked SIs to the working edge nodes. Let $k$ be the number of instances that failed due to an attack on the edge node. In this part, we will perform the k-resilient SIs re-instantiation subject to minimal edge resource usage. The idea of minimizing the edge resource usage instead of delay/latency is important here. The resource and computation capabilities of edge servers are limited and expensive. It is important to utilize them efficiently considering one edge is down due to an attack. Note that this approach is reactive and does not consume resources until the attack takes place. Our third problem of optimal re-instantiation of attacked services is stated as follows, \par 

\textbf{\textit{Problem 3 (Service Recovery Placement (SRP)):}} Given a service placement $x_e^s$ and a failed server, the problem is to find optimal recovery placement locations $z_e^s$ of SIs of each service type $s$ with their resource and delay requirements such that the $z_e^s$ $\neq$ $x_e^s$ and the objective is to minimize the edge resource usage. \par
We will develop the binary ILP-based formulation for solving SP, PSVM, and SRP problems in the following sections.
\subsection{Service Placement (SP)}
\label{Sec:SP}
We formulate the SP problem as an ILP model that enables us to find the optimal choice of edge servers to place the service instances. We use a single objective function to minimize the maximum edge resource usage and service delay, and control the relative importance of resource usage vs. service delay by using a parameter $\alpha$. Most of the vehicular services are latency-sensitive applications and related to driving tasks that make delay as an important objective function for the SP. Thus, minimizing the maximum delay will help to satisfy adequate delay requirements and make service availability faster for the vehicles. The rationale for using resource usage is to efficiently utilize the limited edge resources and have enough room for service demand scale-up as the service demands are dynamic in nature. \par 
The solution to our SP problem is represented by a binary variable $x_e^s$. If edge node $e$ hosts service $s$ then $x_e^s$ is 1. Otherwise, it is 0. Now, the objective function is formulated as:
\begin{equation}
\underset{x}{minmax} \hspace*{3mm} \left(\alpha \sum_{e\in E} \varphi_e + (1-\alpha) \sum_{e\in E} \sum_{s\in S} d_{e}^s \right)x_e^s
\label{eqObj1}
\end{equation}
Here, $\varphi_e$ is the edge resource usage and $d_{e}^s$ is the average service delay observed by vehicles while requesting for service $s$ from edge node $e$. \par 
\textit{Definition 1 (Edge Resource Usage):} We define the edge resource usage as the ratio between the resources that instances of different services will consume and the total available resources at the edge node. We formulate it as:  \par
\begin{equation}
\varphi_e = \frac{\sum_s R_s }{C_e}, \forall e\in E, \forall s\in S,
\label{eq:resouceUsage}
\end{equation}
Here, $C_e$ is the total resource units available at the edge node $e$ and $R_s$ is the resource units required to deploy the instance of service type $s$. \par 
 \par 
\textit{Definition 2 (Service Delay):} In our model, the service delay $d_{e}^s$ is the delay observed by vehicles while accessing service $s$ from edge node $e$, and consists of two components, i.e. propagation delay and queuing delay. As the resources are adequate to handle the service request load, the queuing delay is negligible in normal working condition. \par 
Finally, with the objective in (\ref{eqObj1}), the placement of service is subjective to multiple constraints. \par
\begin{equation}
\sum_{e\in E}x_e^s = I_s, \   \ \forall s\in S
\label{EQ:resilience}
\end{equation}
\begin{equation}
\sum_{s\in S}x_e^s d_{e}^s \le D_s, \   \ \forall e\in E
\label{EQ:delay}
\end{equation}
\begin{equation}
\sum_{e\in E}x_e^s R_s \le C_e, \   \ \forall s\in S
\label{EQ:resource}
\end{equation} 
\begin{equation}
x_e^s \in \{0,1\}; \  \forall s\in S, \forall e\in E
\label{EQ:binaryx}
\end{equation} 
Constraint (\ref{EQ:resilience}) guarantees that from the total required $I_s$ instances, each instance of service type $s$ must be placed on a different edge server node to ensure redundancy and availability of a service from multiple edge servers. This constraint is important in our design as it helps to increase failure tolerance by facilitating service availability of service type $s$ from multiple edge nodes, and eventually, this will help us in our resilient service placement framework. The calculation of  $I_s$ is based on $\mathbb{C}$ and given by:
\begin{equation}
I_s=\ceil[\bigg]{\frac{\lambda_s}{\mathbb{C}}} ; \forall s\in S
\label{eq:instance}
\end{equation}
Constraint (\ref{EQ:delay}) ensures that the service delay experienced by vehicles requesting service $s$ should be less than the maximum delay threshold of that service $D_s$. Constraint (\ref{EQ:resource}) ensures that the available resources at the edge node are not exhausted while deploying instances of different service types. Finally, condition (\ref{EQ:binaryx}) defines the decision variable $x_e^s$ as a binary integer decision variable. 

\subsection{Proactive Secondary V2E Mapping (PSVM)}
In this section, we develop an ILP-based proactive optimization formulation to map the affected vehicles to the attack-free working edge nodes upon an attack on a server which hosts the services for the vehicles. In the IoV applications, where vehicles maintain real-time connections with edge nodes to get assistance in driving, failure in service availability can be hazardous. Therefore, this part of our work focuses on ensuring resilience in service provisioning to the affected vehicles by mapping them to attack-free edge nodes, until the recovery takes place. The mapping of vehicles to different SIs on the attack-free edge nodes is subject to minimal delay. The choice of secondary SIs is critical here since the loads from the affected SIs may congest the secondary SIs resulting in creation of queues and causing the vehicles to experience higher delays. We denote the delay during secondary mapping phase as $\psi_e^s$, which is the delay experienced by vehicles while accessing service $s$ from edge node $e$, and calculated as: \par
\begin{equation}
\psi_e^s = d_{prop} + d_{queue}
\end{equation}
To compute the queuing delay $d_{queue}$ during secondary mapping, we model the edge computation system as an M/D/1 queue, where arrival occurs with $\lambda_s(t)$ according to Markov stochastic model and the service processing rate is deterministic (serving at rate $\mathbb{C}$), as discussed in Section \ref{Sec:ServiceModel}. We assume that there is no waiting queue if,
\begin{equation}
\lambda_s+ \lambda_{i} \le \mathbb{C}
\end{equation}
Here, $\lambda_s$ is the pre-attack traffic availing service from a given SI, and $\lambda_i$ is the attack-affected traffic for the same service type, where $i \in \zeta_{a}$ and $\zeta_{a}$ is the set of SIs under attack. The total must be less than its processing capacity $\mathbb{C}$ for queuing delay to be zero. Otherwise, a queue will be created and the average waiting time for service $s$ over the edge node will be calculated as \cite{queueMD1}:
\begin{equation}
d_{queue}=\frac{\grave{\lambda_s}}{2\mathbb{C}(\mathbb{C}-\grave{\lambda_s})} 
\end{equation}
Here, $\grave{\lambda_s}$ represents the number of vehicles in the queue, and calculated as $(\lambda_s+\lambda_{i})-\mathbb{C}$. Finally, the service delay during secondary mapping is calculated as,
\begin{equation}
\psi_e^s = \frac{1}{|V|}\sum_{v\in V}\frac{dist(v,s)}{c} + \frac{\grave{\lambda_s}}{2\mathbb{C}(\mathbb{C}-\grave{\lambda_s})} 
\end{equation}
Here, $dist(v,s)$ is the euclidean distance between vehicle $v$ and the node where service $s$ is deployed, and $c$ is the propagation speed of the signal through the communication medium. \par 
Minimizing the service delay during recovery phase is the objective of our PSVM problem and is formulated as:
\begin{equation}
\underset{Y}{min} \ \ \  \psi_e^s Y
\label{eqObj2}
\end{equation}
The solution to our objective function is $Y$, and it is defined by two binary variables as,
\begin{equation}
Y= \begin{pmatrix}
y_{e,\kappa_1}^s \\
y_{e,\kappa_2}^s
\end{pmatrix} 
\end{equation}
Here, $y_{e,\kappa_1}^s$ is the candidate-1 server location and $y_{e,\kappa_2}^s$ is the candidate-2 server location for secondary V2E mappings. This is a proactive approach that selects $\kappa_1$ and $\kappa_2$ secondary mappings in advance to avoid or minimize additional delays and service disruptions. The decision variable $y_{e,\kappa_1}^s$ and $y_{e,\kappa_2}^s$ is set to 1 to indicate the mapping of attack-affected vehicles. The following constraints must be ensured to perform optimal V2E mappings.
\begin{equation}
y_{e,\kappa_1}^s + y_{e,\kappa_2}^s\le 1; \  \forall e\in E, \forall s\in S
\label{EQ:mapping2}
\end{equation}
\begin{equation}
y_{e,\kappa_1}^s \le x_e^s; \  \forall e\in E, \forall s\in S
\label{EQ:mapping20}
\end{equation}
\begin{equation}
y_{e,\kappa_2}^s \le x_e^s; \  \forall e\in E, \forall s\in S
\label{EQ:mapping200}
\end{equation}
\begin{equation}
y_{e,\kappa_1}^s \psi_e^s \le y_{e,\kappa_2}^s \psi_e^s; \  \forall e\in E, \forall s\in S
\label{EQ:greaterlesser}
\end{equation}
\begin{equation}
\sum_{e\in E}y_{e,\kappa_1}^s =1, \   \ \forall s\in S
\label{EQ:mapping21}
\end{equation}
\begin{equation}
\sum_{e\in E}y_{e,\kappa_2}^s =1, \   \ \forall s\in S
\label{EQ:mapping22}
\end{equation}
\begin{equation}
\sum_{s\in S}y_{e,\kappa_1}^s \ge 1, \   \ \forall e\in E
\label{EQ:mapping23}
\end{equation}
\begin{equation}
\sum_{s\in S}y_{e,\kappa_2}^s \ge 1, \  \forall e\in E
\label{EQ:mapping24}
\end{equation}
\begin{equation}
\sum_{s\in S}Y \psi_e^s \le D_s, \   \ \forall e\in E
\label{EQ:delay2}
\end{equation}
\begin{equation}
y_{e,\kappa_1}^s,y_{e,\kappa_2}^s \in \{0,1\}, \   \ \forall s\in S, \forall e\in E
\label{EQ:binaryy}
\end{equation} 
Constraint (\ref{EQ:mapping2}) ensures the V2E mappings for two secondary candidates cannot be done to the SIs deployed at the same edge node. Constraint (\ref{EQ:mapping20}) and (\ref{EQ:mapping200}) ensure that V2E mappings for secondary candidates can only be made to SIs placed on an edge node in the SP problem. Constraint (\ref{EQ:greaterlesser}) guarantees that the delay observed over secondary candidate $\kappa_1$ is less than the secondary candidate $\kappa_2$. Constraint (\ref{EQ:mapping21}) and (\ref{EQ:mapping22}) assures that only one SI is selected for each candidate of the secondary mapping. Constraint (\ref{EQ:mapping23}) and (\ref{EQ:mapping24}) guarantee an edge server can host a secondary SI or a set of secondary SIs for candidate $\kappa_1$ and $\kappa_2$ of different service types. Constraint (\ref{EQ:delay2}) ensures that the delay experienced by attack-affected vehicles requesting service $s$ should be less than the allowable delay threshold $D_s$. Finally, constraint (\ref{EQ:binaryy}) defines the decision variables as binary variables. 

\subsection{Service Recovery Placement (SRP)}
In this section, we formulate the ILP model of optimal service recovery placement for re-instantiation of SIs under attack. The recovery placement of SIs aims to minimize the average edge resource usage assuming that one edge node is down due to an attack. The rationale for using edge resource usage as an objective is important for efficient utilization of limited edge resources and decreasing the possibility of switching-on new edge servers. The inefficient distribution and resource usage will also limit the ability of the EN in satisfying increasing future demands. Hence, the objective of our proposed SRP framework is to minimize the post-attack total edge resource usage, and formulated as,
\begin{equation}
\underset{z}{min_{e \in E}} \hspace*{3mm} \sum_{s\in S} \left(\frac{\delta_s}{\breve{C_e}} \right)  z_e^s
\label{eqObj3}
\end{equation}
Here, $\delta_s$ is the resource demand for service recovery and $\breve{C_e}$ is the residual capacity of edge resources after primary service placements. The solution to SRP is defined by a binary variable $z_e^s$. If service $s$ is placed at edge node $e$, $z_e^s$ is 1. Otherwise, it is 0. The recovery placement is subject to following constraints. 
\begin{equation}
x_e^s \le 1 - z_e^s , \   \ \forall e\in E \ \forall s\in S
\label{EQ:mapping30}
\end{equation}
\begin{equation}
\sum_{e\in E}z_e^s = I_s, \   \ \forall s\in S
\label{EQ:mapping3}
\end{equation}
\begin{equation}
\sum_{s\in S}z_e^s d_{e}^s \le D_s, \   \ \forall e\in E
\label{EQ:delay3}
\end{equation}
\begin{equation}
\sum_{e\in E}z_e^s \left(\delta_s + R_s \right)\le C_e, \   \ \forall s\in S 
\label{EQ:resource3}
\end{equation} 
\begin{equation}
z_e^s \in \{0,1\}; \  \forall s\in S, \forall e\in E
\label{EQ:binaryz}
\end{equation} 
Constraint (\ref{EQ:mapping30}) is a resilient placement constraint that guarantees that SRP instance locations must be different from SP locations of \textit{Problem 1}. Constraint (\ref{EQ:mapping3}) specifies that each instance of service type $s$ must have a recovery placement location and be deployed onto a different edge node. Constraint (\ref{EQ:delay3}) is on the fulfilment of the delay threshold requirement. Constraint (\ref{EQ:resource3}) guarantees that the resource capacity of edge nodes is not exceeded. Finally, condition (\ref{EQ:binaryz}) defines $z_e^s$ as a binary decision variable.

\section{Resilient Service Placement Framework: Architecture and Algorithm}
\label{Sec:ResilientPlacementFramework}
\begin{figure}[htbp]
	\centering
	\includegraphics[width=3in, height=1.45in]{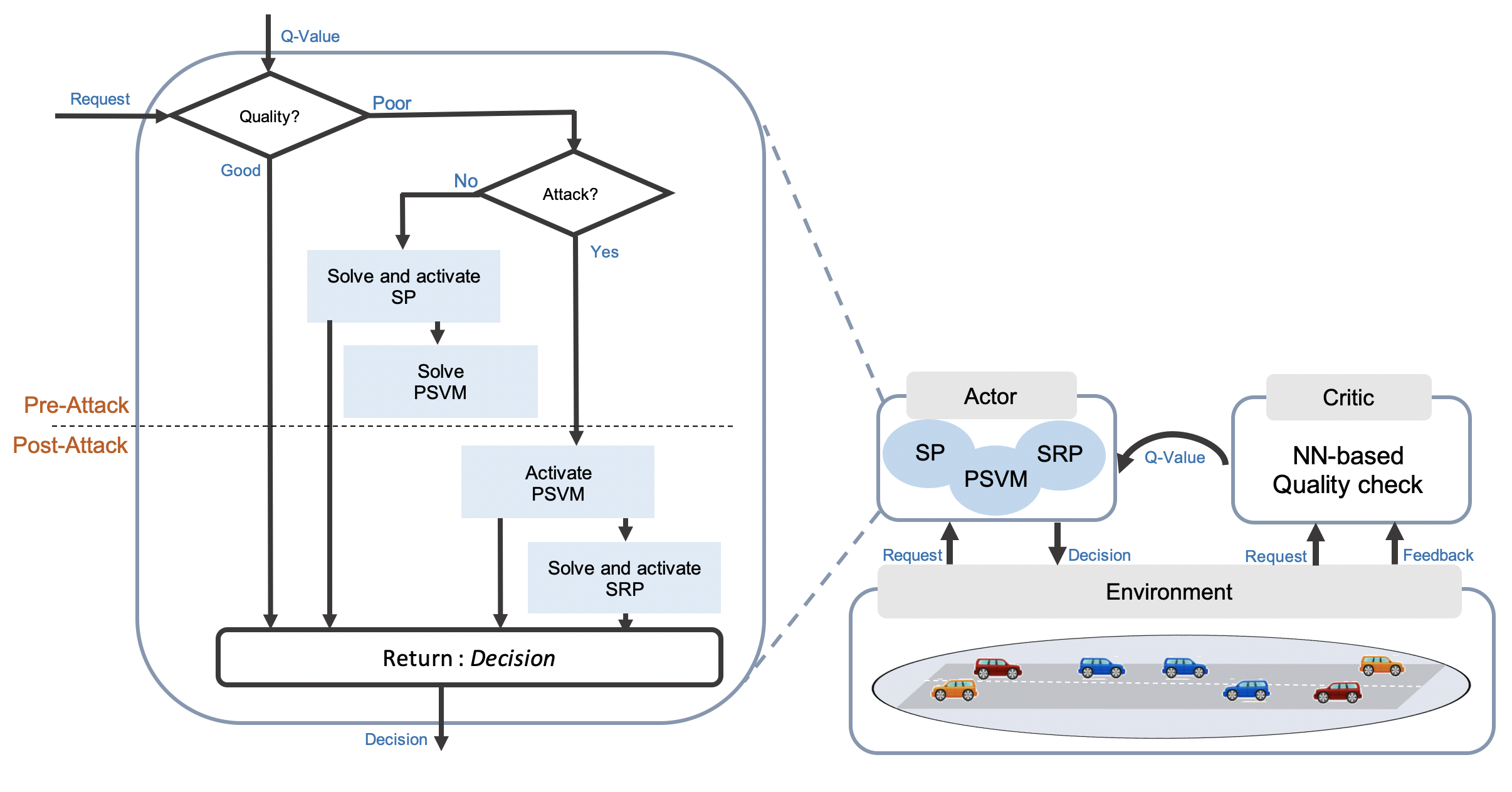}
	\caption{The architecture of proposed resilient service placement framework}
	\label{fig:framework}
\end{figure}

In the previous section, we develop three optimal ILP models for SP, PSVM and SRP. We elaborate our idea to solve these formulated problems in this section and describe our proposed resilient service placement framework. We depict the architecture of our proposed resilient service placement framework in Fig. \ref{fig:framework}. \par 
The key concept of our service placement framework is based on the DRL framework. A static deployment that fixes servers for hosting services is not effective for IoV considering the mobile nature of vehicles and dynamics in the requests for different types of services. It is therefore imperative that the real-time environment be taken into consideration while mapping a service to an edge server. Therefore, we exploit the actor-critic DRL model with ILP formulations to solve the resilient service placement framework. \par 
An actor is a primary function which generates action by using the policy function. Whereas the critic estimates a value function, we call as quality-value (i.e. $Q_{value}$), needed to maintain good performance in the network by dynamically re-optimizing the actor policy formulations. The $Q_{value}$ changes between 0 to 1, a small value implies poor performance necessitating re-optimization. An actor needs to select actions with the maximum quality value, i.e $\mathfrak{a}$ = arg max $Q_{value}$. Here, $\mathfrak{a}$ denotes the action. The design of the critic model is similar to that in our work \cite{DRLD-SP}. In our proposed design, the critic network is a neural network which collects feedback and the request for which feedback was generated from the environment, as shown in Fig. \ref{fig:framework}. From the user perspective, our objective is to minimize the service delay observed from vehicles in accessing service $s$ from the associated edge server $e$. Therefore, the feedback function is modeled as:
\begin{equation}
\mathcal{F}=\mathbb{E}\left[d_e^s(t)\right]
\label{eq:reward}
\end{equation}
The feedback is a response, an agent receives by the environment for the corresponding action. The critic network updates its parameters $\theta$ to minimize the mean square loss function $\mathcal{L}_Q$ based on the feedback and it's corresponding request parameters. The loss function is computed as:
\begin{equation}
\mathcal{L}_Q(\theta)=\frac{1}{\mathcal{N}}\sum_{i=1}^{\mathcal{N}}\left[(y_{t_i}-Q_{value}(\mathfrak{a};\theta))^2\right]
\label{eq:loss}
\end{equation}
Here, $y_t$ is a target value which is calculated as:
\begin{equation}
y_t =
\begin{cases}
\sigma(D_s,\mathcal{F}) & \mathcal{F} < D_s\\
0 & \text{else}
\end{cases} 
\label{eq:yt}
\end{equation}
Where $\sigma(D_s,\mathcal{F})$ is the standard deviation between delay threshold and feedback. The higher the deviation, the better the model is in terms of delay. \par
In our actor design, we want to ensure resilience by using three ILP formulations, i.e. SP, PSVM, and SRP. Here, SP is the solution for the placement of services in the normal working state (i.e. \textbf{\textit{attack-free}} state). One of the important constraints for the SP solution is the redundancy (Eq (\ref{EQ:resilience})) where we limit the deployment to no more than one instance of each service type on the same edge node. This is to ensure the availability of the same type of service from multiple edge nodes. Along with SP, we also solve PSVM in the attack-free state. The PSVM is the proactive approach and determines an effective secondary V2E mapping for vehicles to be used when the network is under attack. It plays a major role in our algorithm and helps avoid service disconnection or breakage and minimizes the average performance loss for the sensitive service types, such as driving. \par  
Upon an \textbf{\textit{attack}}, our framework promptly activates PSVM secondary mappings to maintain service availability to the attack-affected vehicles and minimize the impact of the attack until recovery takes place. In the meantime, we start to solve the SRP to find the best possible service instance recovery placements for the given state of the environment. Once the SRP is solved, it is activated to enhance the network performance in terms of delay and the network will become resilient to the attack with a negligible loss in the network performance, which we verify in detail in Section \ref{Sec:PerformanceStudy}. The SRP decision will remain in force until the attacked edge is restored and be ready again to participate in the service placement. \par

\begin{algorithm}
	\DontPrintSemicolon
	\KwInput{service profile, edge profile, critic NN-model}
	
	\For{t=1,2,3,....}
	{	
		Observe the service request as 4-tuple input \\
		Calculate $I_s$ for all $s\in S$ using (\ref{eq:instance}) \\
		\If{t==1}
		{
			Solve \textit{SP} using (\ref{eqObj1}) for an initial solution $\overline{x_e^s}$ \\ 
			Solve \textit{PSVM} using (\ref{eqObj2}) for an initial solution $\overline{y_{e,\kappa_1}^s}$ and $\overline{y_{e,\kappa_2}^s}$\\
			Set Decision = $\overline{x_e^s}$\\
		}
	\Else
		{
			Find new \textbf{\textit{$x_e^s$}} from $Actor(request,Q_{value})$ \\
		}
		Obtain feedback at \textit{Critic}  \\
		Calculate $Q_{value}$ from NN-model \\
		Observe $Q_{value}$ at $Actor(request,Q_{value})$ \\
	}
	\KwOutput{Decision on service location}
	\caption{Resilient Service Placement}
	\label{Alg:main}
\end{algorithm}

\begin{algorithm}
	\DontPrintSemicolon
	\KwInput{request and $Q_{value}$}
	
	Observe the service request as 4-tuple input \\
	\If{$Q_{value}$=Good}
	{
		Find $x_e^s$ from last saved \textit{SP} solution \\
		Set decision = $x_e^s$ \\
	}
	\If{$Q_{value}$=Poor}
	{
		\If{Attack=FALSE}{
			Solve \textit{SP} using (\ref{eqObj1}) for $x_e^s$ \\ 
			Solve \textit{PSVM} using (\ref{eqObj2}) for $y_{e,\kappa_1}^s$ and $y_{e,\kappa_2}^s$\\
			Decision = $x_e^s$ \\
		}
		\If{Attack=TRUE}
		{
			Find edge-node under attack as $\mathcal{N}_a$ \\
			Find set of SIs under attack as $\zeta_{a}$ \\
			Find set of vehicles influenced by attack as $\mathcal{V'}$ \\
			Set Decision = $y_e^s$ for $\mathcal{V'}$ \\
			Solve \textit{SRP} using (\ref{eqObj3}) for $z_e^s$\\ 
			Find recovery placements as $z_e^s(\zeta_{a})$ from $z_e^s$ for $\zeta_{a}$  \\
			Set $x_e^s(\mathcal{N}_a)=null$ \\
			Update Decision = $x_e^s$ $\oplus$ $z_e^s(\zeta_{a})$ \\ 
		}
	}
	\KwReturn{Decision on service location}
	\caption{$Actor(request,Q_{value})$}
	\label{Alg:actor}
\end{algorithm}

Our proposed attack-resilient service placement framework is solved iteratively to take into account of the dynamic nature of the IoV network and described in Algorithm \ref{Alg:main} and Algorithm \ref{Alg:actor}. Algorithm \ref{Alg:main} serves as a general resilient service placement framework and works recursively from steps 1-12. First, it initializes the service profile, edge profile, and the critic neural network (NN) model. Then, it starts with the data collection and SI calculation in step 2 and 3, respectively. The data collection is the request for set of services $S$ by vehicles $V$ following the service model discussed in Section \ref{Sec:ServiceModel}. Step 4-7 indicate that in the beginning at time $t=1$ where SP and PSVM are solved for initial decisions. The service placements from SP are sent back to the environment as $\overline{x_e^s}$. For the later iterations, step 9 calculates a new solution for $x_e^s$ using Algorithm \ref{Alg:actor}. \par 
The decision is communicated back to the environment with the purpose that vehicles start getting services from the edge servers where their desired service type is deployed. At each time unit $t$, the vehicles generate feedback and send it to the critic model in step 10. The feedback indicates the performance of an environment in terms of the delay observed by the vehicles. In step 11, the critic NN model uses feedback as an input feature to generate the Q-value. This value is sent to the actor model for further evaluation and to find a new placement decision. \par  
Algorithm \ref{Alg:actor} computes the resilience-based actor policy to find new optimal placements for services. Step 1 indicates the reception of input (service request) at time unit $t$. Steps 2-4 indicates good-quality system performance where vehicles continue to receive service from the same edge nodes. On the contrary, if the system performance quality is poor, the network will check for an attack or no-attack scenario in steps 5-17. The type of attack we study here is discussed in Section \ref{Sec:ProblemDefinition} where the edge node fails because of any type of outage attack which results in the blackout of the edge node. Step 6-9 indicate the no-attack case, but due to poor $Q_{value}$, the SP and PSVM are re-optimized to find new placements and secondary V2E mappings. \par 
Step 10-17 indicate the failure of an edge node and the state where network is under attack. Step 11 finds out the node under attack as $\mathcal{N}_a$. Steps 12-13 find out the set of service instances under attack as $\zeta_{a}$ and the set of vehicles affected by the attack as $\mathcal{V'}$. In step 14, the secondary V2E mappings solved from PSVM are communicated to $\mathcal{V'}$ to minimize the impact of the attack until recovery takes place. Steps 15-16 solve the SRP to find the recovery placements for $\zeta_{a}$. Step 17 excludes node $\mathcal{N}_a$ from the placement decision. Finally, the decision is updated in step 18 by including instance recovery placements into it. \par  

\section{Performance Study}
\label{Sec:PerformanceStudy}
In this section, we present the performance study of our optimization models and proposed framework. We first start with a summary of experimental settings and then evaluate our framework for several primary performance metrics. We also provide insights into the effectiveness of different IoV environments with varying vehicle densities and compare it against an existing baseline solution of BR reservation.\par

\subsection{Experimental settings}
\label{Sec:ExperimentalSettingss}
The IoV environment and vehicle trajectories used in our evaluation for generating service requests are from real-world vehicle mobility dataset, provided by crawdad \cite{sanfrancisco}. In this set of the environment, the data is generated from a maximum of 500 taxis travelling the city of San Francisco. From the big city given, we extract an area of 15x15 $km^2$ for use in our experiments. In addition to the San Francisco dataset, we evaluate our model with two other real-world datasets provided for the city of Rome \cite{rome} and Beijing \cite{beijing1,beijing2}. This is to validate the effectiveness of our models for varying traffic volumes and dynamics in the vehicle localities in different city environments. We extract the same size of the area (i.e. 15x15 $km^2$) from each dataset for fair comparison in the experiments. The vehicle trajectories are generated from a maximum of 194 taxis and 500 taxis in the city of Rome and Beijing, respectively. The choice of the above datasets is significant, as all are urban environments with different traffic densities and varying the number of active taxis at a given time slot. \par
\begin{figure}[hbt!]
	\captionsetup[subfigure]{justification=centering}
	\centering
	\begin{subfigure}{.15\textwidth}
		\centering
		\includegraphics[width=1.2in,height=1in]{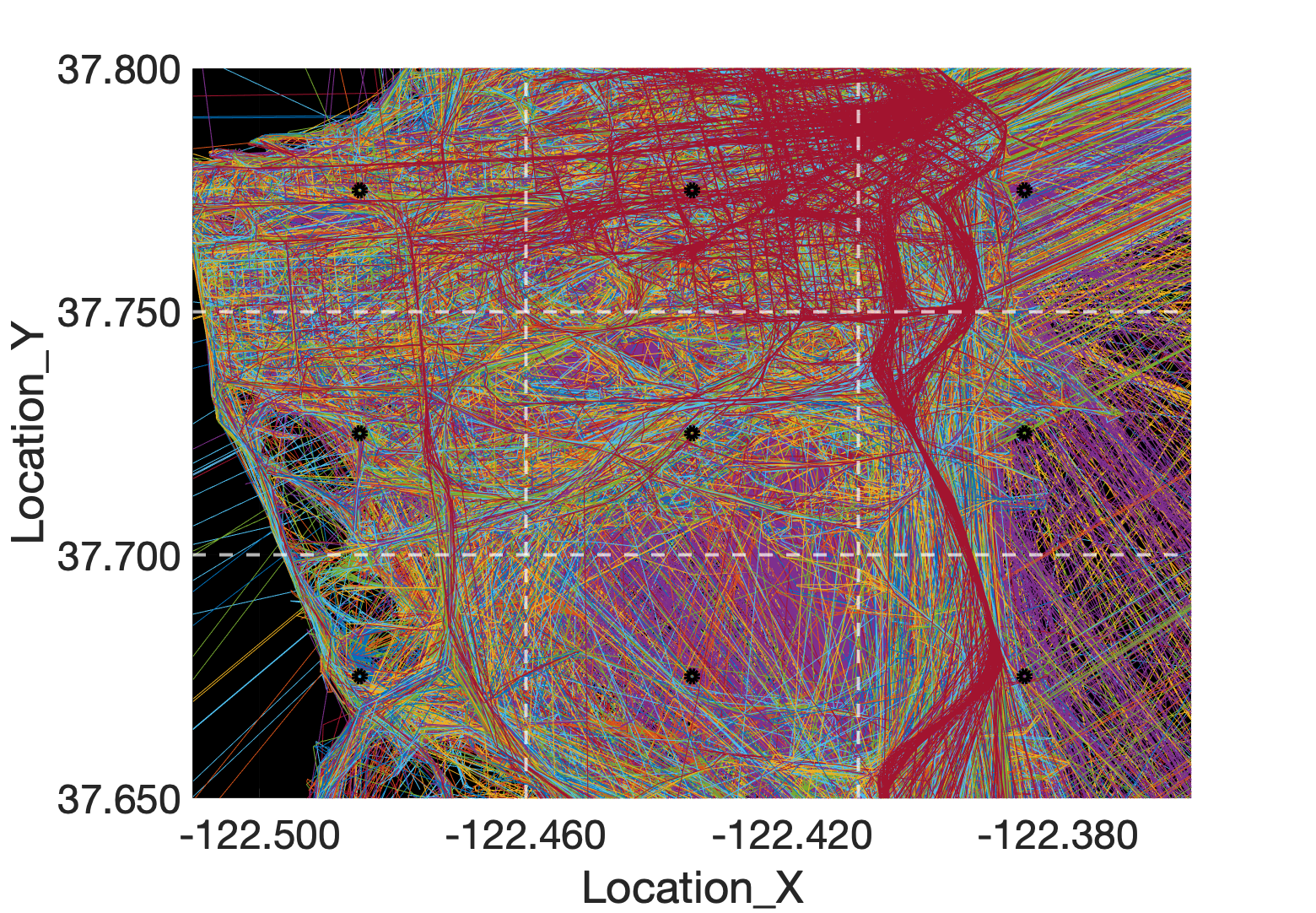}  
		\caption{San Francisco}
		\label{fig:data1}
	\end{subfigure}
	\begin{subfigure}{.15\textwidth}
		\centering
		\includegraphics[width=1.2in,height=1in]{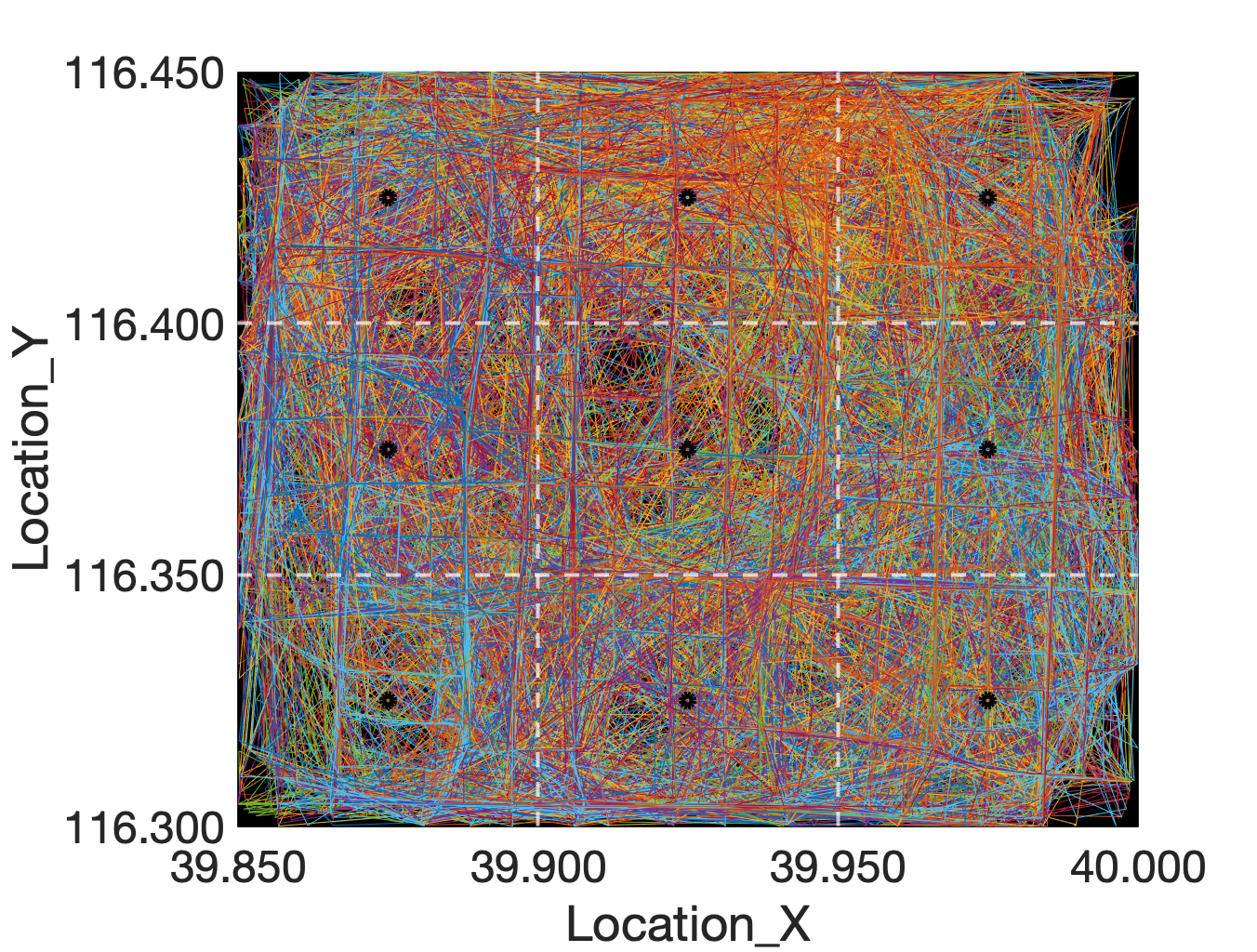}  
		\caption{Beijing}
		\label{fig:data2}
	\end{subfigure} 
	\begin{subfigure}{.15\textwidth}
		\centering
		\includegraphics[width=1.2in,height=1in]{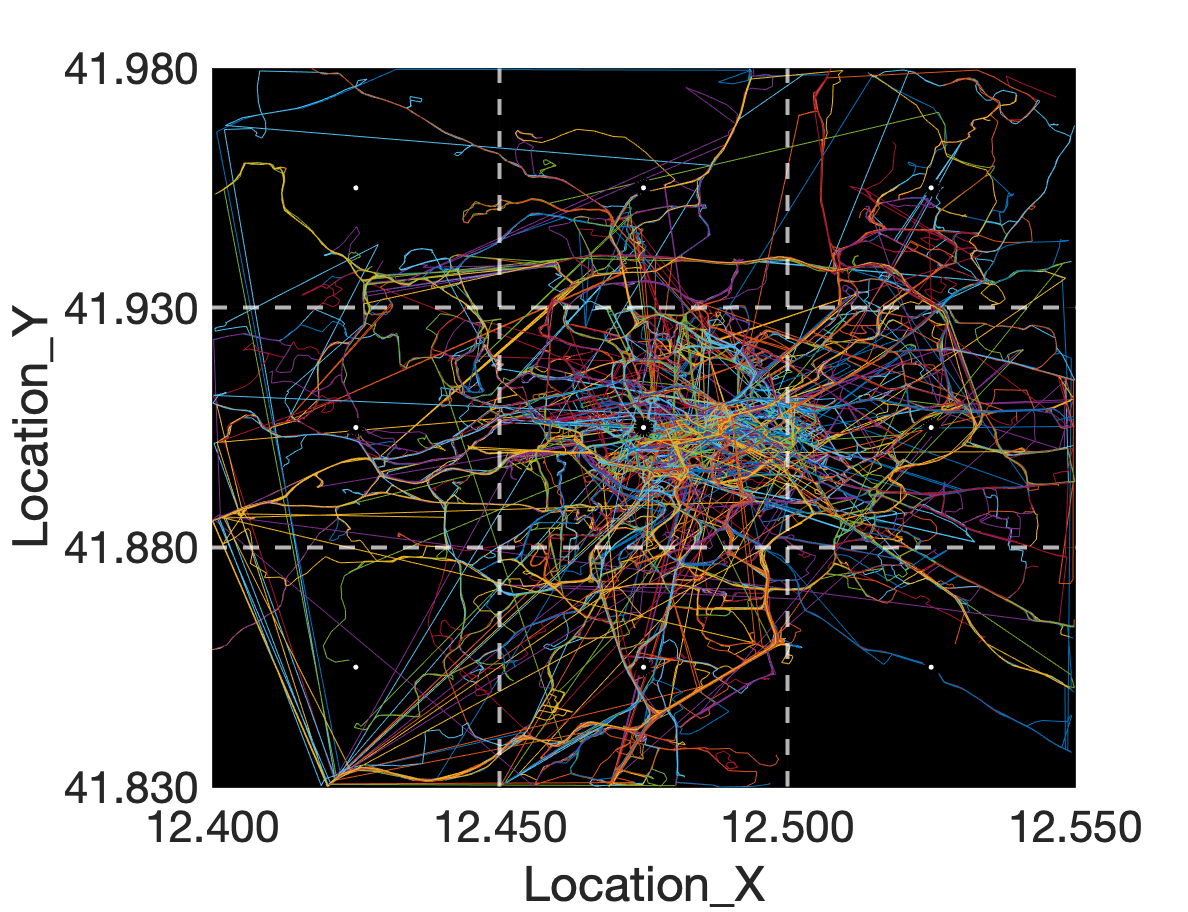}  
		\caption{Rome}
		\label{fig:data3}
	\end{subfigure}
	\caption{Vehicle trajectories for different city environments}
	\label{fig:dataset}
\end{figure}
The implementation of our proposed ILP formulations and DRL framework is carried out using MATLAB software. We consider an edge system with $9$ areas shown in Fig. \ref{fig:dataset}, and each area has an edge node $e \in E$ with an abstract measure of resources as $C_e=100$ units. The system has ability to provide $8$ different types of services where placement of single instance of service type $s$ requires $R_s=\{10, 12, 14, ... 24\}$ resource unit, and has delay threshold requirements of $D_s(ms)=\{50, 60, 70, ..., 120\}$. Each instance has a processing capacity of $\mathbb{C}=30$. For the SP problem, we use $\alpha=0.5$ to give a fair importance to both resource usage and delay. The placement of the total number of SIs (i.e. $I_s$) varies with time $t$ and depends on the arrival rate $\lambda_s(t)$, and calculated using (\ref{eq:instance}). \par 
For the critic neural network design, we conduct a comprehensive experimental study to find the best hyperparameters. We use a fully-connected feed-forward network with 4 hidden layers, each with 512, 256, 64, and 32 neurons, respectively. We use a hyperbolic tangent sigmoid for the activation of hidden layers, and the output layer is a single neuron that expresses the $Q_{value}$ with the linear transfer function for the activation. To avoid overfitting, the learning rate of 0.01 is used to train the network. The maximum number of episodes to train a network is 1500 with each episode having a maximum of 20 iterations and a batch size of 100. The parameters of the critic network are updated every 5 time slots. All the experiments are evaluated on a system with Intel Corei5 2GHz and 8GB RAM.

\subsection{Results}
In this section, we plot and discuss results for two different network states, i.e. \textbf{\textit{pre-attack}} and \textbf{\textit{post-attack}}. When only the SP model is active during an attack-free state, we call it as Pre-Attack SP (\textbf{\textit{PrA-SP}}). On the contrary, the PSVM is operative when the network is under-attack, and SRP assists with improving network performance by recovering the failed SIs. In this regard, we formally indicate the network performance when the PSVM and SRP is active as Post-Attack PSVM (\textbf{\textit{PoA-PSVM}}) and Post-Attack SRP (\textbf{\textit{PoA-SRP}}).

\subsubsection{Performance of Proposed Framework}
In this section, we verify the effectiveness of our proposed attack-resilient framework using different performance metrics. \par
\begin{figure}[hbt!]
	\captionsetup[subfigure]{justification=centering}
	\centering
	\begin{subfigure}{.21\textwidth}
		\captionsetup[subfigure]{justification=centering}
		\centering
		\includegraphics[width=1.7in,height=1.2in]{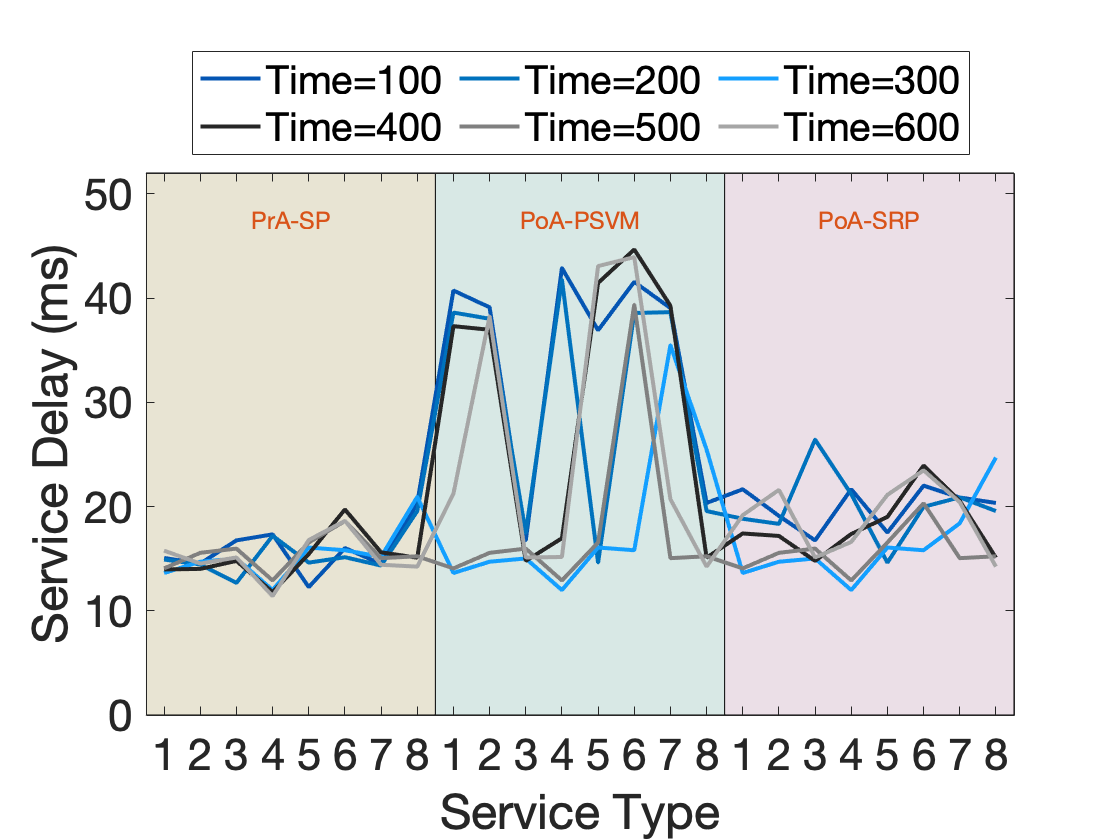}  
		\caption{}
		\label{fig:delay1SF}
	\end{subfigure}
	\begin{subfigure}{.21\textwidth}
		\centering
		\includegraphics[width=1.7in,height=1.2in]{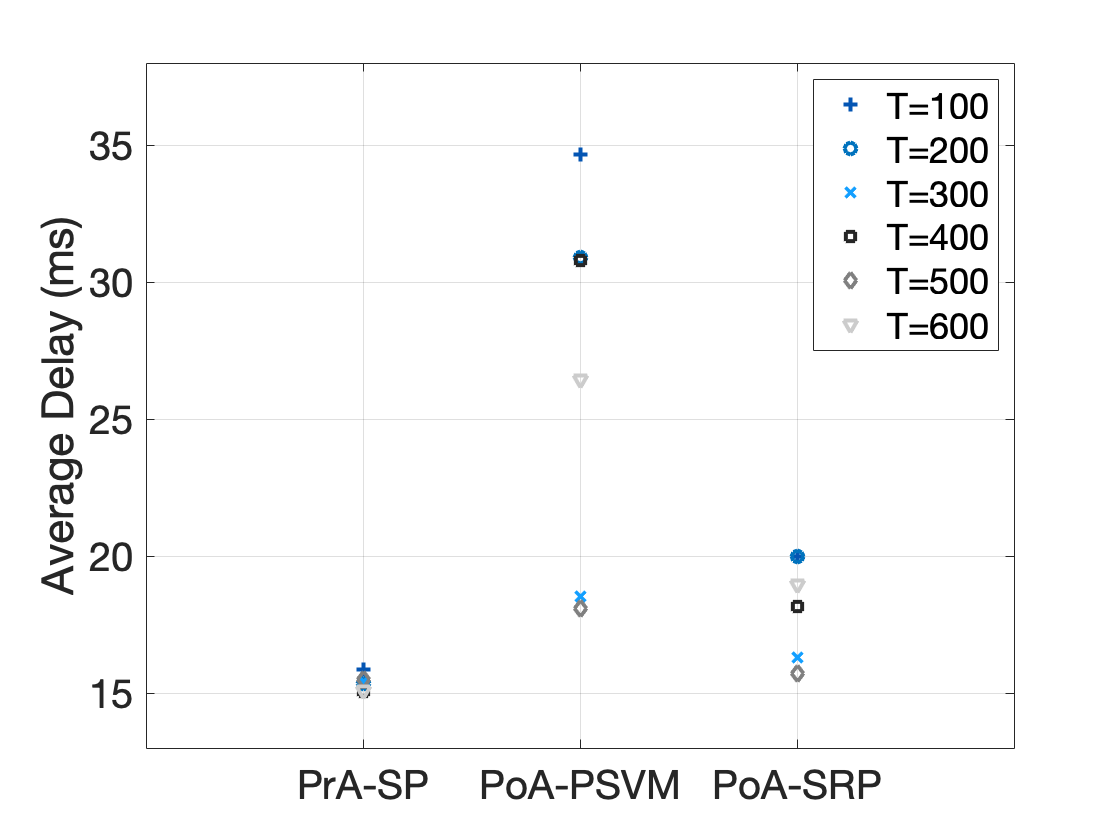}  
		\caption{}
		\label{fig:delay2SF}
	\end{subfigure} 
	\caption{Delay Performance}
	\label{fig:delay}
\end{figure}

 
First, we use the delay/latency metric and plot service delay experienced by vehicles for availing different services before and after an attack. Fig. \ref{fig:delay} shows the service delay for different states of the network when different models are operational. It can be observed in Fig. \ref{fig:delay1SF} that service delay is lowest for SP during the pre-attack state. Once the network is under attack and upon the failure of an edge node, our framework quickly activates the PSVM mapping to avoid any service disruption and maintain resilience. However, the delay observed by the vehicles during the PoA-PSVM state is higher due to the addition of affected vehicles to the attack-free edge nodes, which can increase the load beyond their capacity and add to queuing delay which will eventually increase the service delay. The PSVM mapping is temporary, and once the recovery placement is complete in PoA-SRP, the service availability is again faster in our framework with shorter service delay observed by vehicles. \par
To further analyze the delay performance, Fig. \ref{fig:delay2SF} illustrates the average delay for all services for different time units. It can be seen that average delay is not always high when the PoA-PSVM is active. This is from the fact that the SIs which are failed at the node under attack are observing different traffic volumes at different time units. The busy times may result in higher queuing delays compared to less busy times and eventually, the delay during PSVM mapping tends to vary with time. We can notice that the average delay during T=300 and T=500 is lower and nearly the same as of SRP. \par
\begin{figure}[hbt!]
	\captionsetup[subfigure]{justification=centering}
	\centering
	\begin{subfigure}{.15\textwidth}
		\centering
		\includegraphics[width=1.2in,height=1in]{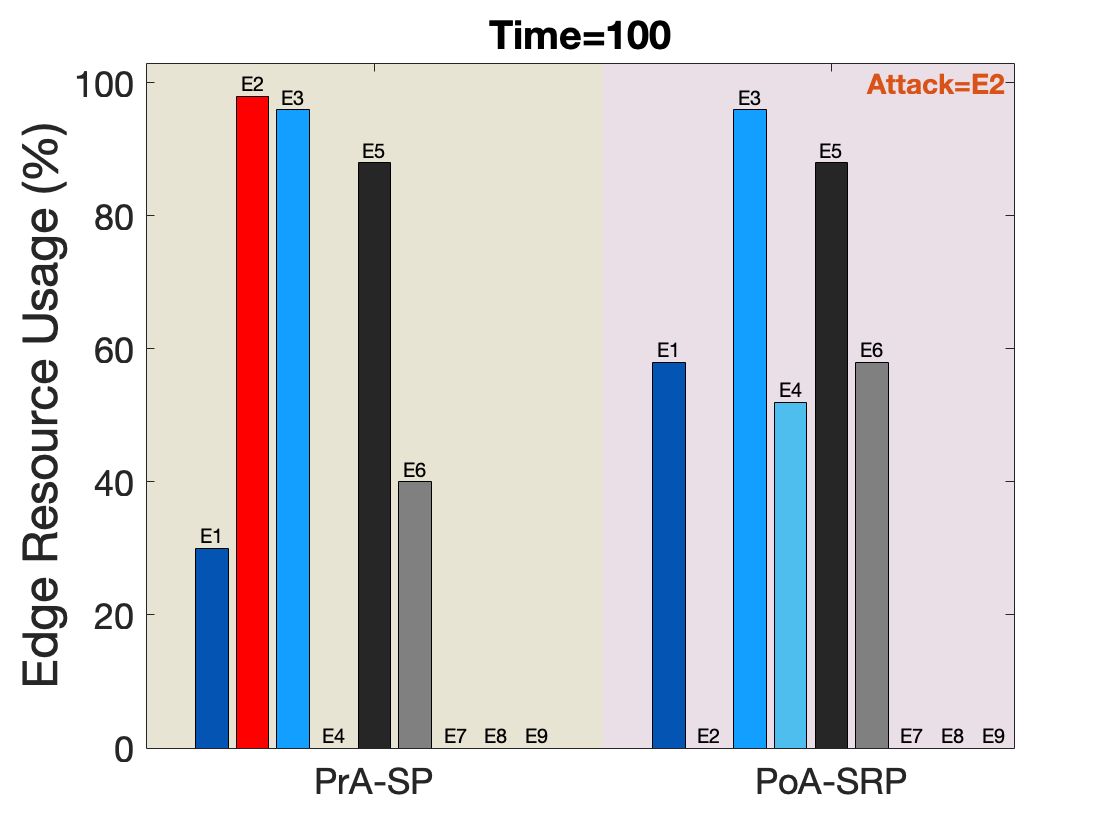}  
		\label{fig:ERU1-SF}
	\end{subfigure}
	\begin{subfigure}{.15\textwidth}
		\centering
		\includegraphics[width=1.2in,height=1in]{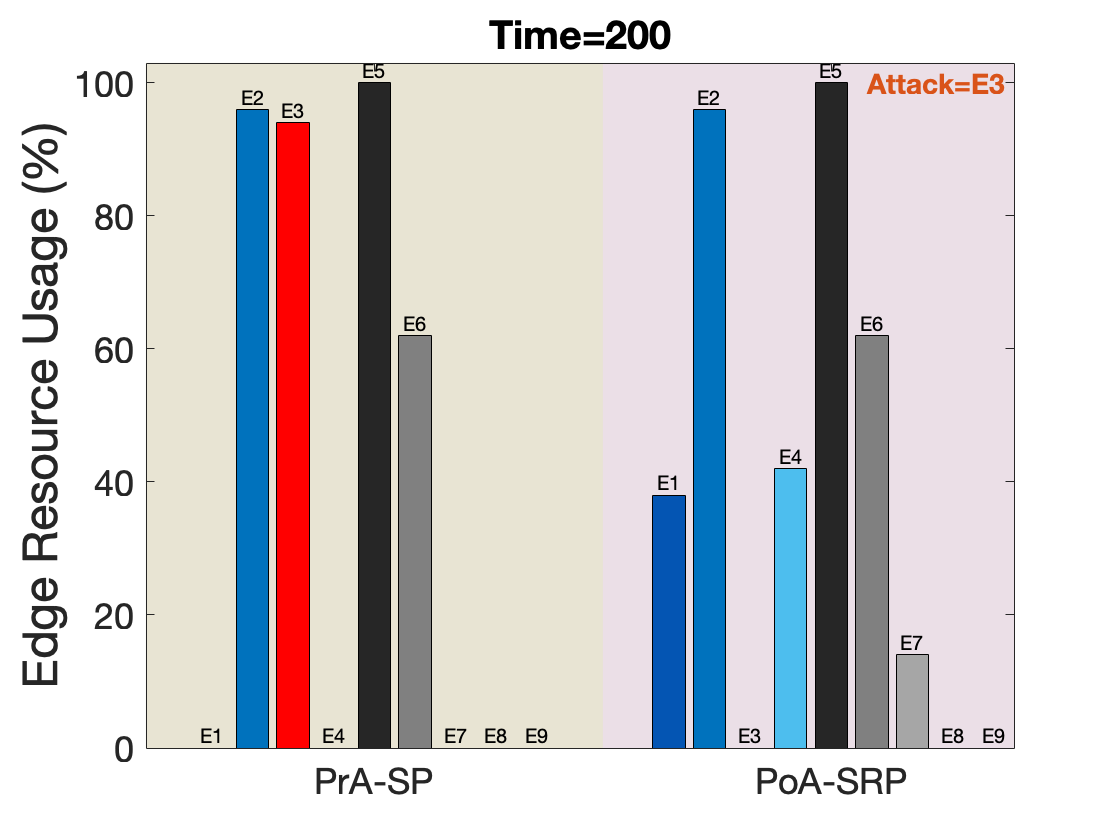}  
		\label{fig:ERU2-SF}
	\end{subfigure} 
	\begin{subfigure}{.15\textwidth}
		\centering
		\includegraphics[width=1.2in,height=1in]{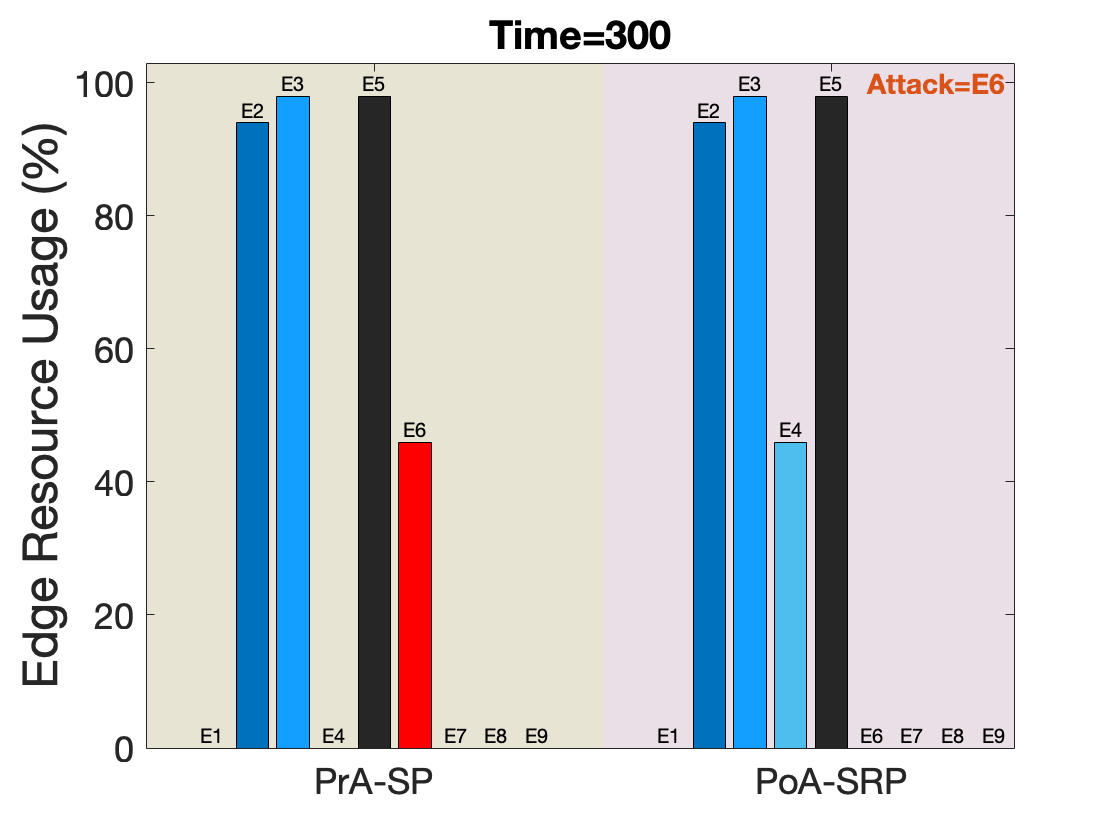}  
		\label{fig:ERU3-SF}
	\end{subfigure} \\
	\begin{subfigure}{.15\textwidth}
		\centering
		\includegraphics[width=1.2in,height=1in]{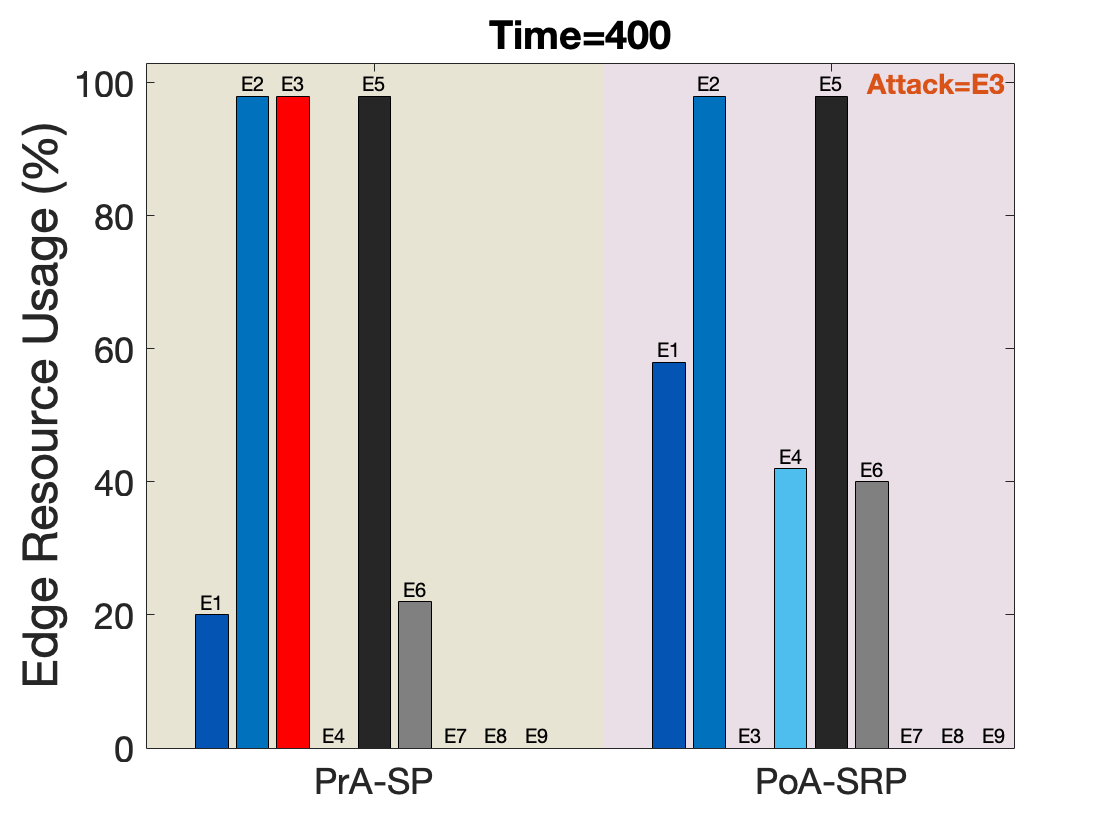}  
		\label{fig:ERU4-SF}
	\end{subfigure} 
	\begin{subfigure}{.15\textwidth}
		\centering
		\includegraphics[width=1.2in,height=1in]{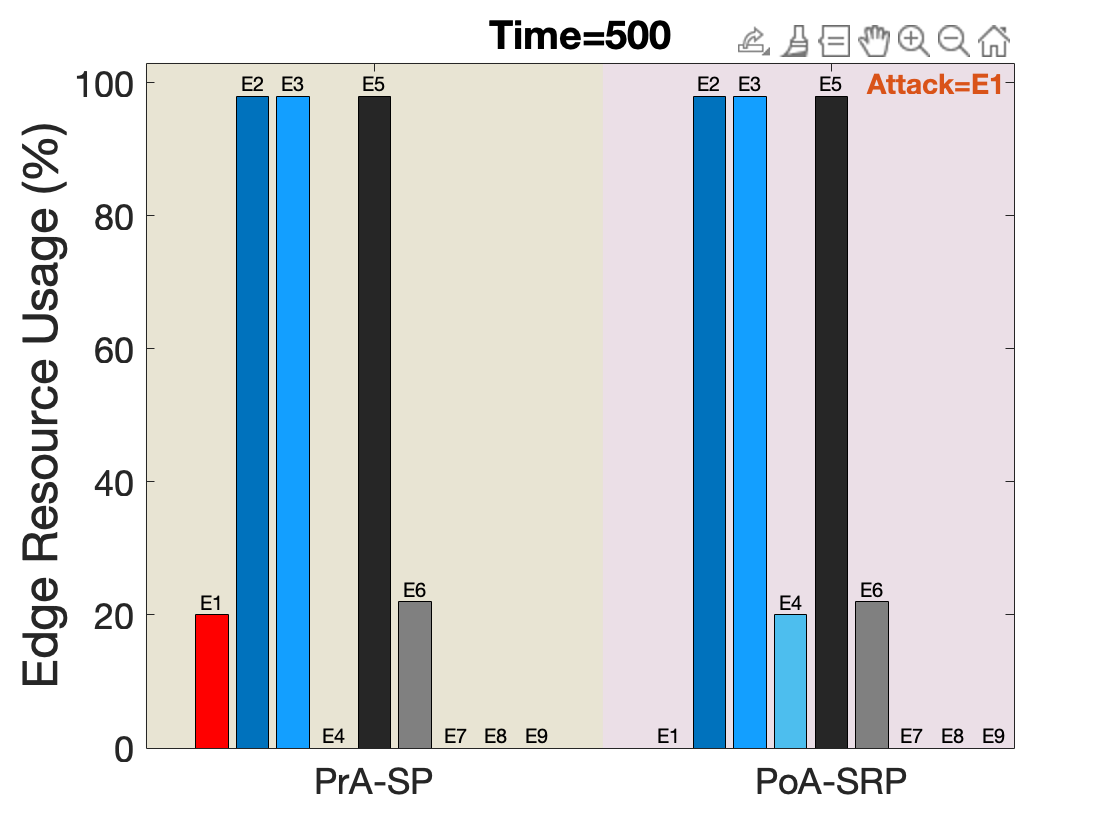}  
		\label{fig:ERU5-SF}
	\end{subfigure}
	\begin{subfigure}{.15\textwidth}
		\centering
		\includegraphics[width=1.2in,height=1in]{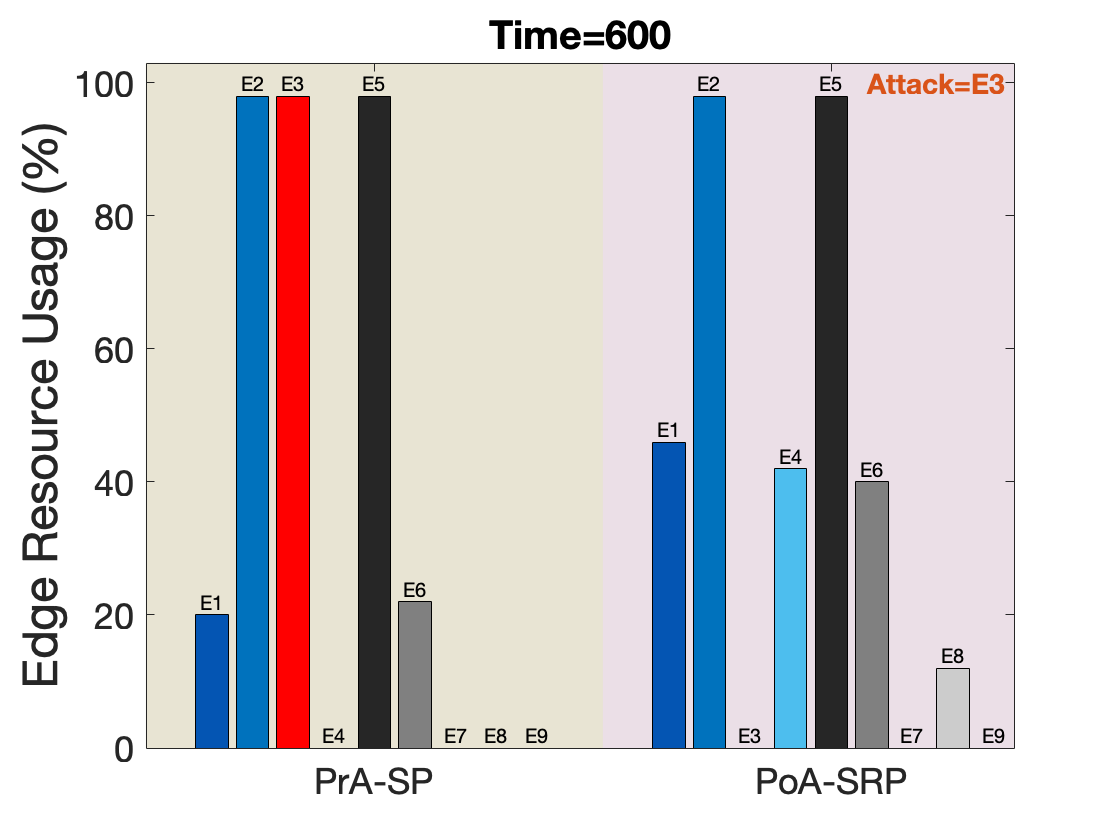}  
		\label{fig:ERU6-SF}
	\end{subfigure} 
	\caption{Edge resource usage}
	\label{fig:ERU-SF}
\end{figure}

 
Fig. \ref{fig:ERU-SF} plots the edge resource usage as the percentage of the ratio between the resources consumed by SIs and the available resources at the edge node. We illustrate the spread of service resources for PrA-SP and PoA-SRP against different nodes under attack at different time units. The red coloured bar on PrA-SP highlights the node under attack. In our framework, the choice of the node under attack is random and demonstrates the case of the heavily-loaded node under attack (at Time=100, 200, 400, and 600), the moderately-loaded node under attack (at Time=300), and finally, the lightly-loaded node under attack (at Time=500). With the objective of minimizing edge resource usage, our proposed algorithm exhibits good performance for all the cases in both pre-attack and post-attack scenarios. The low delay service requirement is satisfied while maintaining less spread of services among different edge nodes. Note that most of the time our proposed solution tries to place and fit SIs into the available active edge nodes instead of powering-on a new edge node. The lesser the spread is, the lesser will be the maintenance cost associated with its functioning. This can be attributed to the fact that for facilitating the same number of requests with fewer edge nodes will lead to lesser energy expenditure and less wastage of resources. \par
Further, in Table \ref{tab:ActiveServers-SF}, we compare the number of edge nodes that are active to facilitate vehicles at different time units for PrA-SP and PoA-SRP scenarios. We observe that on an average our SP model makes 52\% of edge nodes available to the vehicles whereas SRP makes 56\% of edge nodes available throughout the network. This validates that our proposed resilient framework performs close to the attack-free scenario in the event of an attack on a node. Use of fewer servers implies lower cost associated with the servers and better chance to satisfy future demands.\par
\begin{table}[htbp]
  \centering
  \caption{Number of active edge nodes}
  \scriptsize
  \tabcolsep=0.09cm
    \begin{tabular}{lcccccc}
    \toprule
          & \textbf{T=100} & \textbf{T=200} & \textbf{T=300} & \textbf{T=400} & \textbf{T=500} & \textbf{T=600} \\
    \midrule
    \textbf{PrA-SP} & 5     & 4     & 4     & 5     & 5     & 5 \\
    \midrule
    \textbf{PoA-SRP} & 5     & 5     & 4     & 5     & 5     & 6 \\
    \bottomrule
    \end{tabular}%
  \label{tab:ActiveServers-SF}%
\end{table}%

\subsubsection{Comparison with the Baseline Algorithm}
As a baseline for comparison, we use the backup resource (BR) reservation method to assess the performance of our proposed framework. The BR method is a well-known approach in the recent literature to maintain resilience and handle service failure, as discussed in Section \ref{Sec:RelatedWork}. In the BR method, redundancy is introduced by placing one extra instance with resource reservation for each service type to guarantee survivability and service continuity in th event of a failure. \par
\begin{figure}[hbt!]
	\captionsetup[subfigure]{justification=centering}
	\centering
	\begin{subfigure}{.15\textwidth}
		\centering
		\includegraphics[width=1.2in,height=1in]{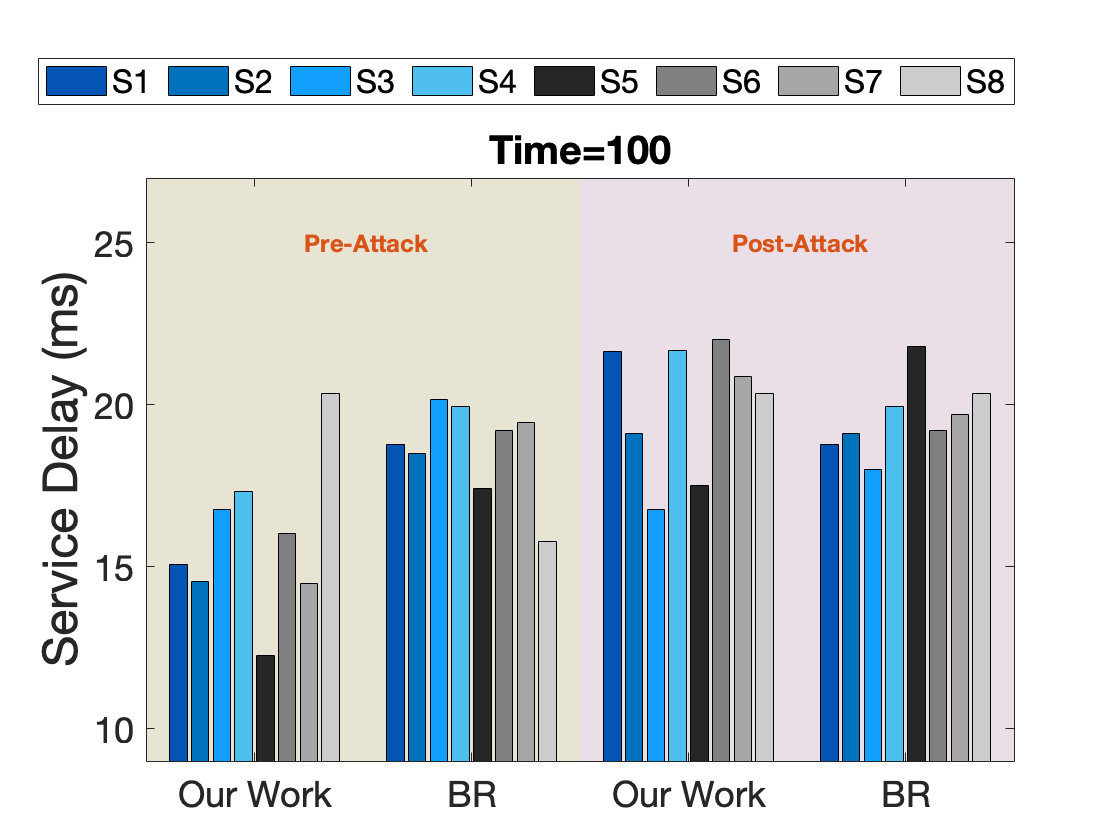}  
		\label{fig:Comp1-SF}
	\end{subfigure}
	\begin{subfigure}{.15\textwidth}
		\centering
		\includegraphics[width=1.2in,height=1in]{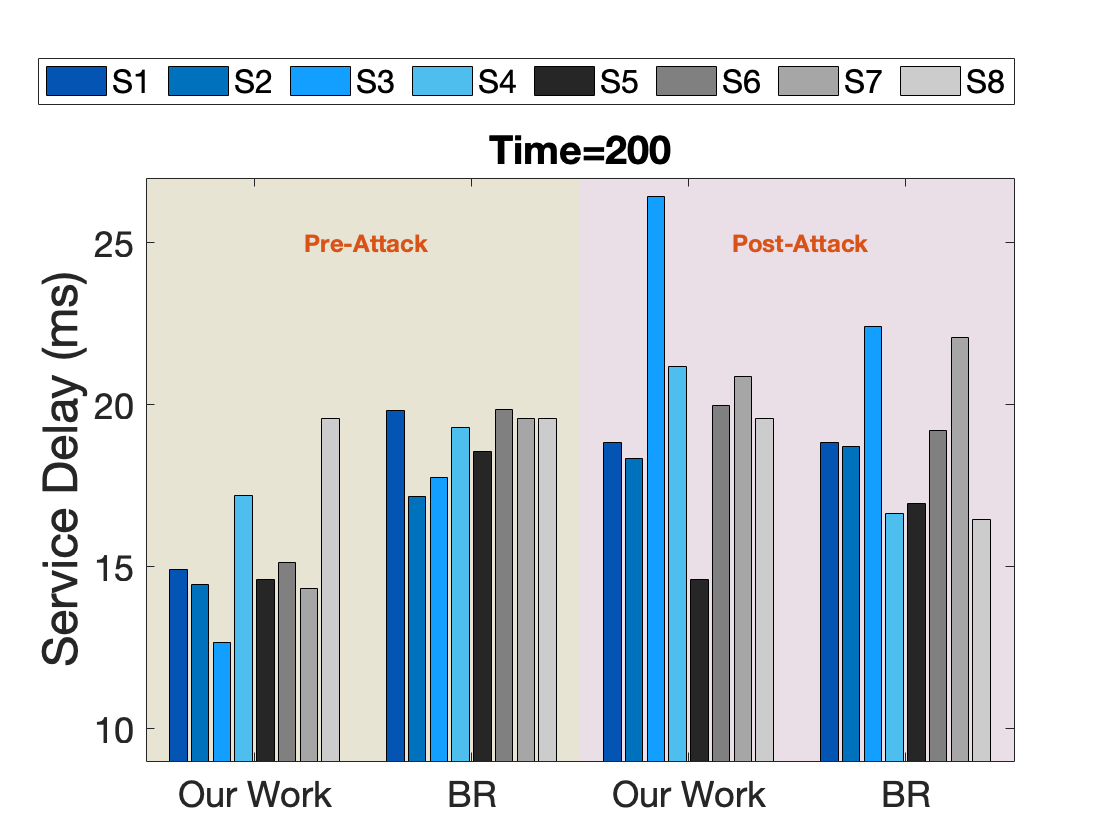}  
		\label{fig:Comp2-SF}
	\end{subfigure} 
	\begin{subfigure}{.15\textwidth}
		\centering
		\includegraphics[width=1.2in,height=1in]{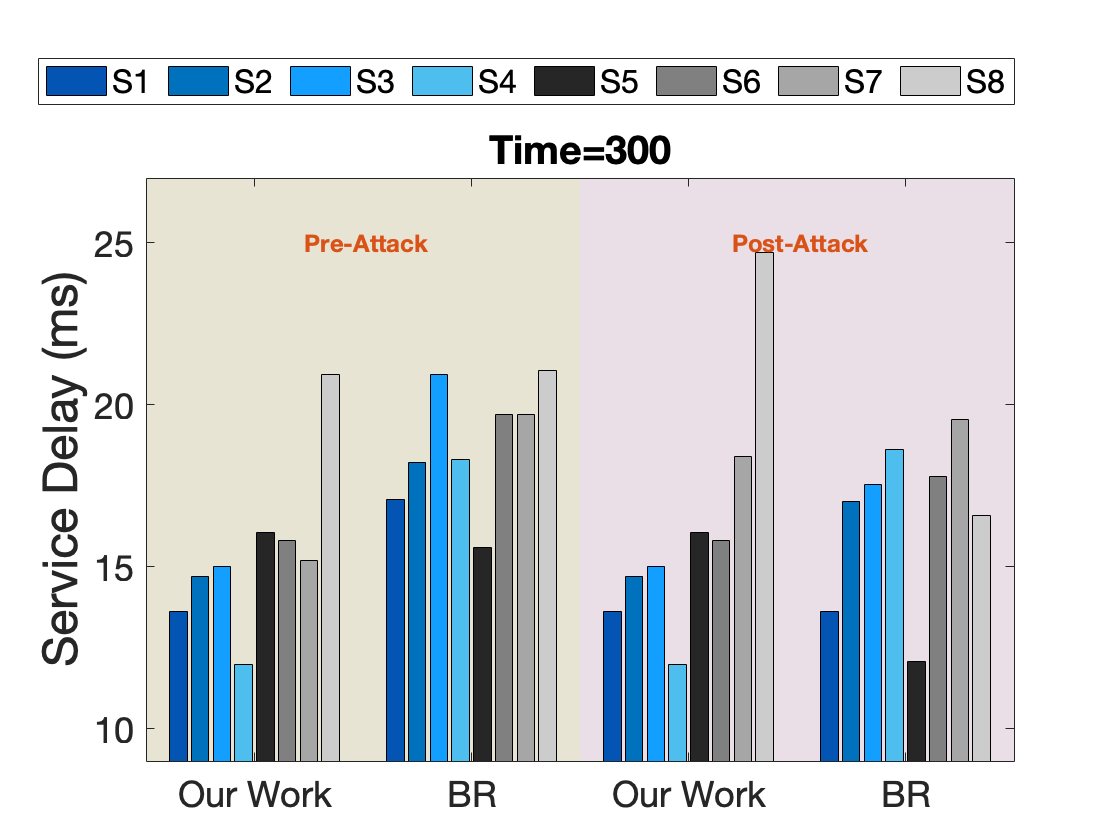}  
		\label{fig:Comp3-SF}
	\end{subfigure} \\
	\begin{subfigure}{.15\textwidth}
		\centering
		\includegraphics[width=1.2in,height=1in]{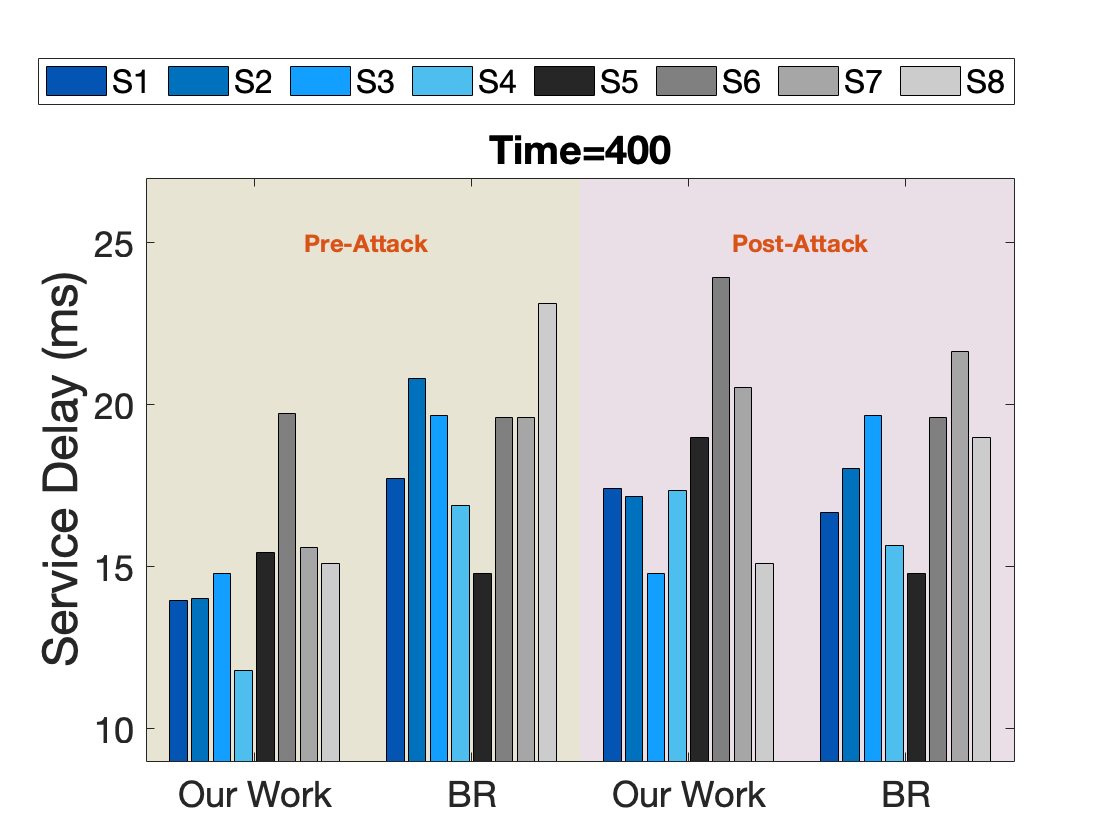}  
		\label{fig:Comp4-SF}
	\end{subfigure} 
	\begin{subfigure}{.15\textwidth}
		\centering
		\includegraphics[width=1.2in,height=1in]{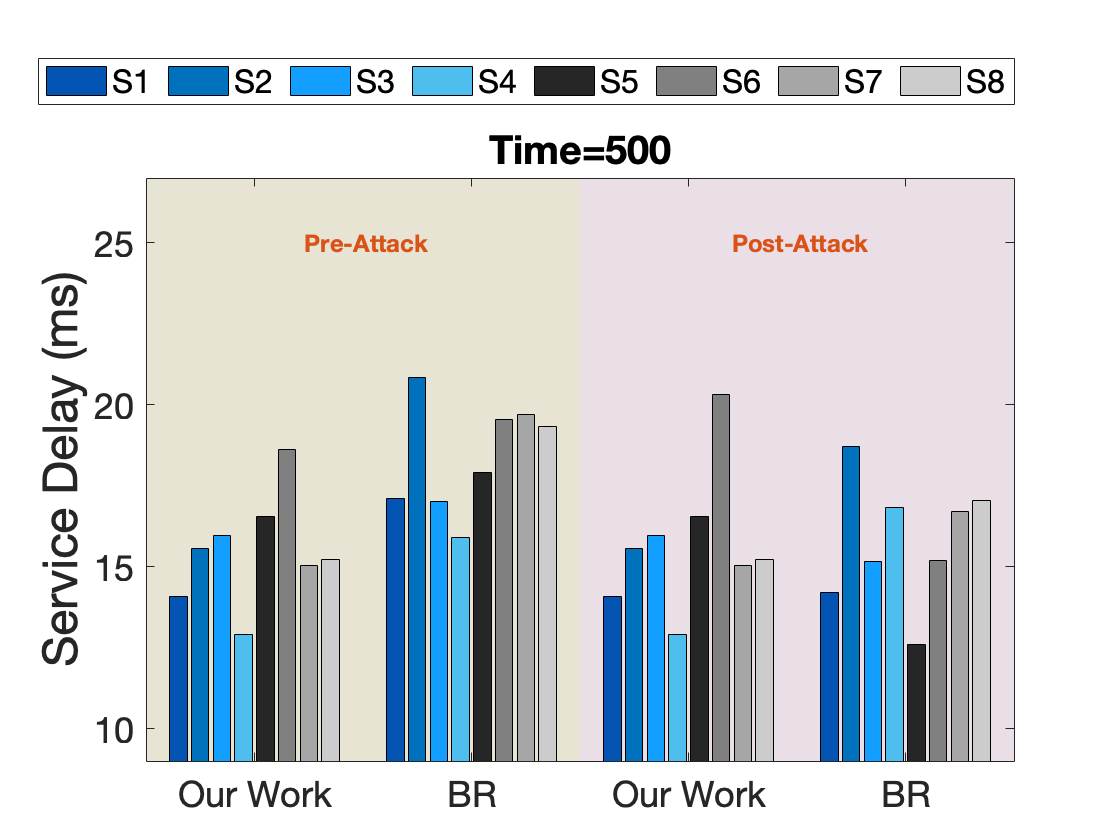}  
		\label{fig:Comp5-SF}
	\end{subfigure}
	\begin{subfigure}{.15\textwidth}
		\centering
		\includegraphics[width=1.2in,height=1in]{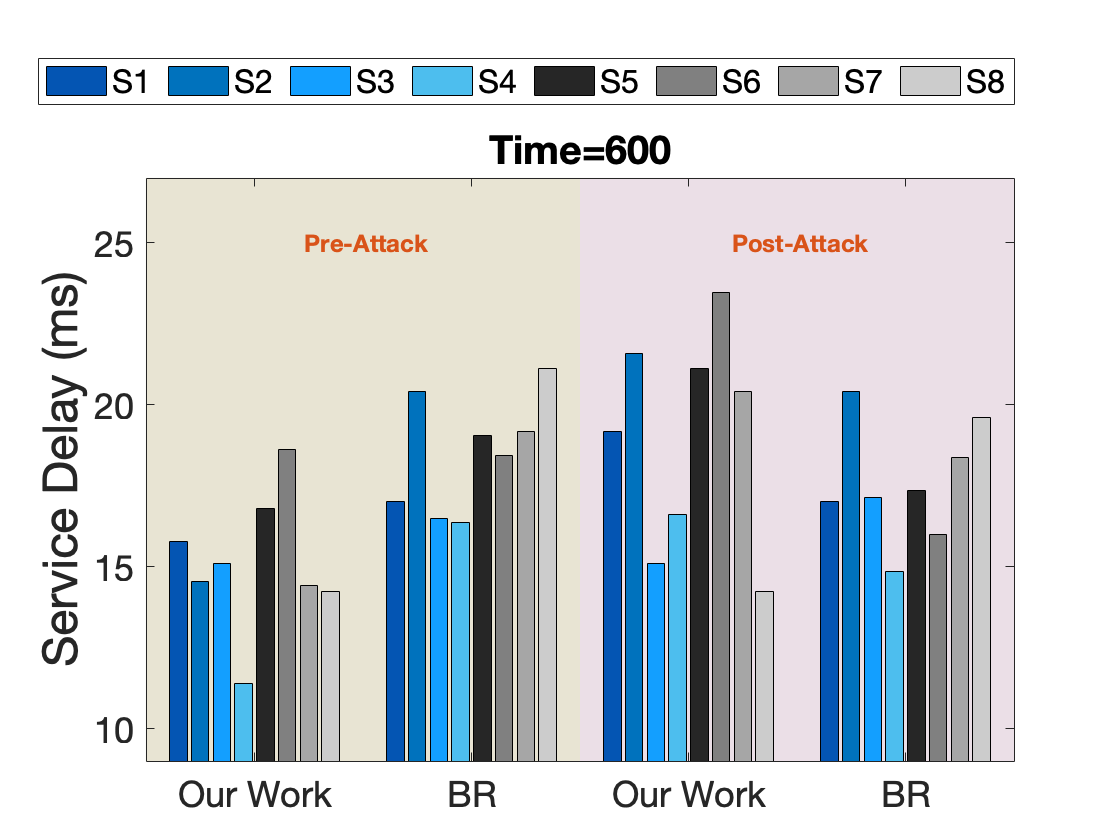}  
		\label{fig:Comp6-SF}
	\end{subfigure} 
	\caption{Comparison of service delay in our framework with baseline}
	\label{fig:Comp-SF}
\end{figure}

 
Fig. \ref{fig:Comp-SF} plots the service delay experienced by vehicles for each service type using our framework and BR method for pre-attack and post-attack scenarios. We first note that in a pre-attack phase where a redundant instance is not active yet, our framework outperforms the BR method. It is from the fact that pre-occupancy of resources to deploy redundant instances will minimize the choice of available resources for the placement of services. Thus, the limitation in the scope of choice will force instances to be placed on less optimal edge nodes and increases the delay experienced by vehicles in accessing the service. On the contrary, the delay experienced in BR for the post-attack scenario is fairly similar to our work but BR consumes more resources and quickly exhausts the limited and expensive edge resources. \par

\begin{figure}[hbt!]
	\captionsetup[subfigure]{justification=centering}
	\centering
	\begin{subfigure}{.23\textwidth}
		\centering
		\includegraphics[width=1.8in,height=1.3in]{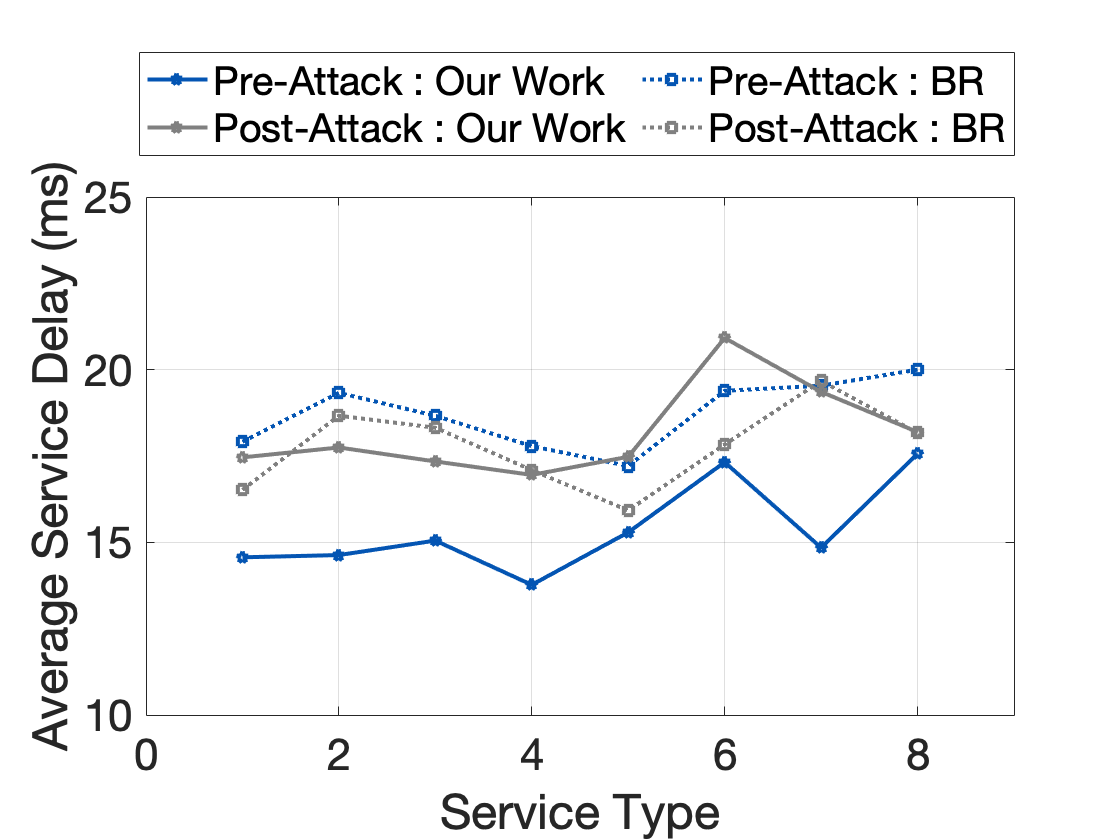}  
		\caption{}
		\label{fig:CompAvg}
	\end{subfigure}
	\begin{subfigure}{.23\textwidth}
		\centering
		\includegraphics[width=1.8in,height=1.3in]{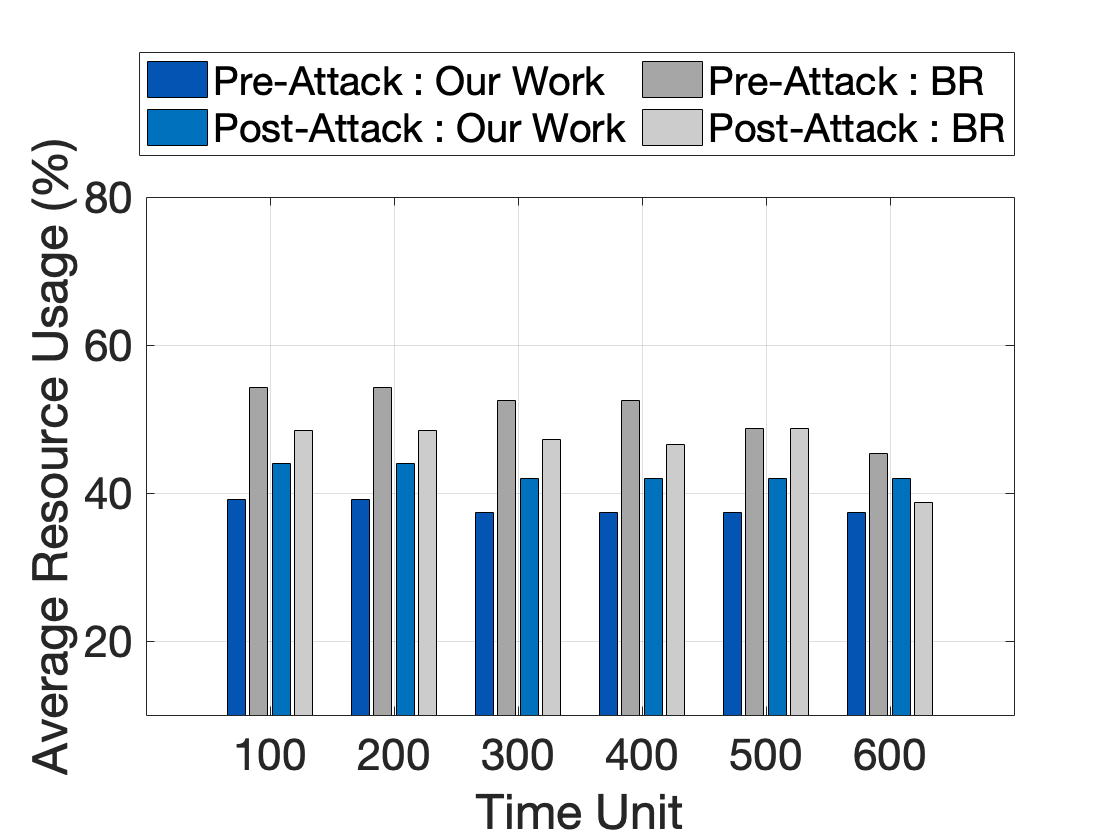}  
		\caption{}
		\label{fig:CompERU}
	\end{subfigure} 
	\caption{Average performance comparison of our framework with baseline}
	\label{fig:delaySFComp}
\end{figure}

Fig. \ref{fig:CompAvg} plots the average delay observed by the vehicles in our method and the BR method to further emphasize the delay difference. In Fig. \ref{fig:CompERU}, we compare the average resource usage by our method and the BR method. As can be observed from the figure, our proposed framework intends to utilize edge resources more effectively accommodating the same demand. Moreover, for BR, the average usage of resources for both pre-attack and post-attack scenarios is typically higher. \par

In Table \ref{tab:RunTime-SF}, we present the run time, in seconds, of different ILP algorithms on a system with Intel Corei5 2GHz and 8GB RAM. We note that the run time for BR and PrA-SP is more or less similar. In contrast, it can be observed that although the proposed algorithm uses three ILP formulations to ensure resilience in service availability, the run time for PoA-PSVM and PoA-SRP is 1/10th of the time observed in PrA-SP and BR. It is because the smaller the space of the feasible solutions, the lesser the run time for the ILP model. We also note from Table \ref{tab:RunTime-SF} that the run time of all the three proposed algorithms put together, is only slightly higher than that of BR.\par

\begin{table}[htbp]
  \centering
  \caption{Run Time}
  \scriptsize
  \tabcolsep=0.11cm
    \begin{tabular}{p{4.915em}cccc}
    \toprule
    \multirow{2}[4]{*}{} & \multicolumn{1}{c}{\multirow{2}[4]{*}{\textbf{BR}}} & \multicolumn{3}{c}{\textbf{Our Method}} \\
\cmidrule{3-5}    \multicolumn{1}{l}{} &       & \multicolumn{1}{c}{\textbf{PrA-SP}} & \multicolumn{1}{c}{\textbf{PoA-PSVM}} & \multicolumn{1}{c}{\textbf{PoA-SRP}} \\
    \midrule
    \textbf{T=100} & 0.1299 & 0.1880 & 0.0222 & 0.0343 \\
    \midrule
    \textbf{T=200} & 0.1957 & 0.1805 & 0.0243 & 0.0344 \\
    \midrule
    \textbf{T=300} & 0.1802 & 0.1689 & 0.0213 & 0.0394 \\
    \midrule
    \textbf{T=400} & 0.1616 & 0.1713 & 0.0196 & 0.0226 \\
    \midrule
    \textbf{T=500} & 0.1824 & 0.1713 & 0.0242 & 0.0308 \\
    \midrule
    \textbf{T=600} & 0.1816 & 0.1713 & 0.0235 & 0.0156 \\
    \bottomrule
    \end{tabular}%
  \label{tab:RunTime-SF}%
\end{table}%

\subsubsection{Impact of different datasets}
In addition to the San Francisco vehicle mobility dataset, we evaluate our framework with two other real-world vehicle mobility datasets (i.e. Rome and Beijing), as discussed in Section \ref{Sec:ExperimentalSettingss}. This is to study and verify the effectiveness of our framework against varying traffic volumes and different geographic locations of different vehicular environments. \par 
\begin{figure}[hbt!]
	\captionsetup[subfigure]{justification=centering}
	\centering
	\begin{subfigure}{.21\textwidth}
		\centering
		\includegraphics[width=1.7in,height=1.2in]{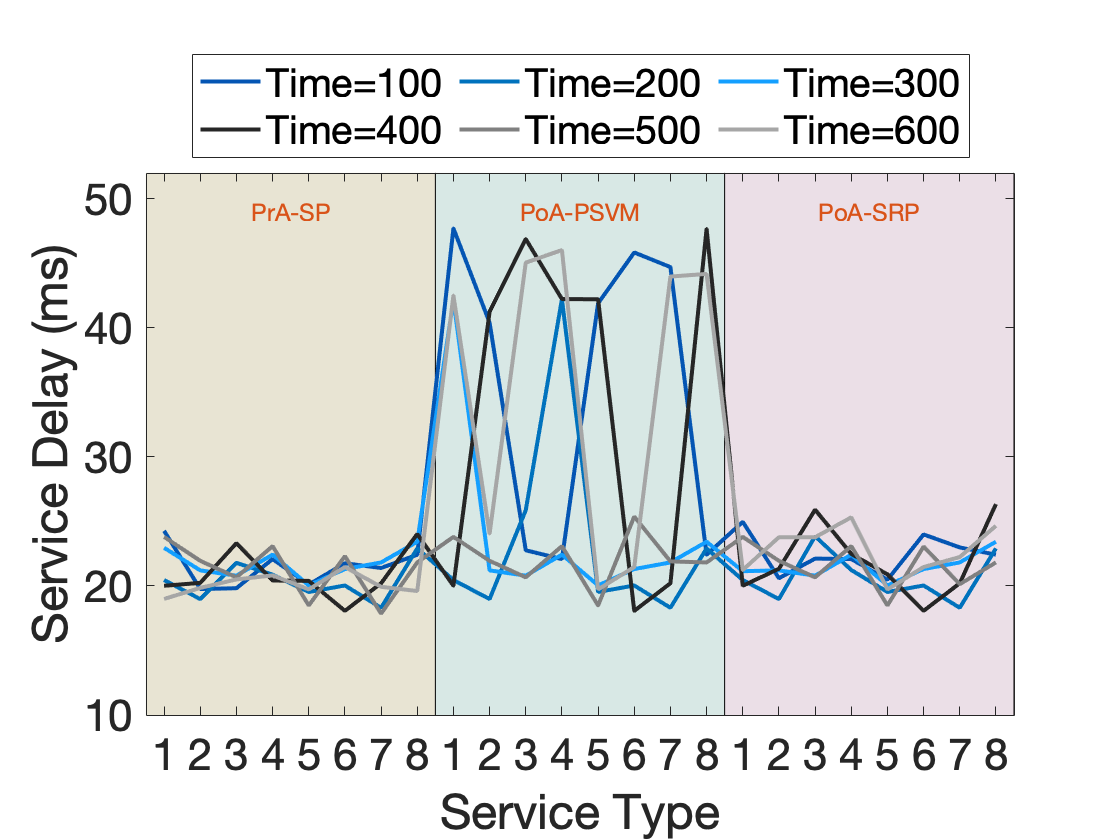}  
		\caption{Beijing}
		\label{fig:delay1BJ}
	\end{subfigure}
	\begin{subfigure}{.21\textwidth}
		\centering
		\includegraphics[width=1.7in,height=1.2in]{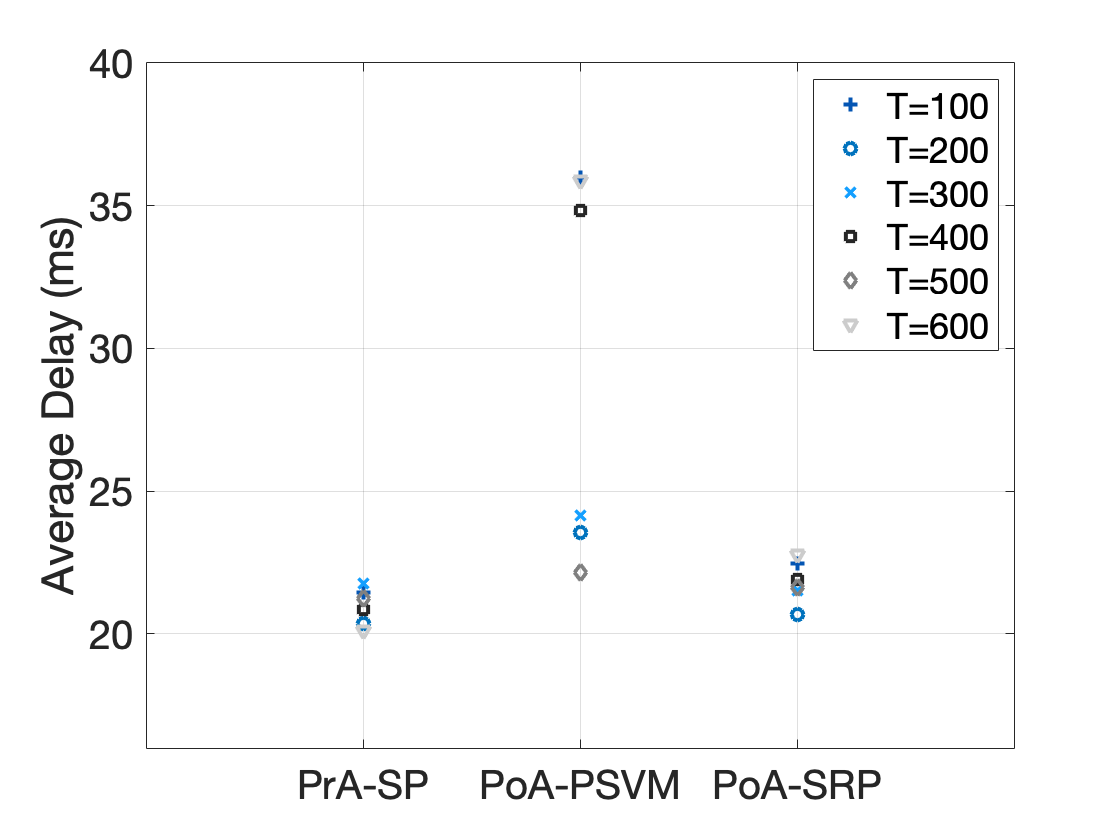}  
		\caption{Beijing}
		\label{fig:delay2BJ}
	\end{subfigure} \\
	\begin{subfigure}{.21\textwidth}
		\centering
		\includegraphics[width=1.7in,height=1.2in]{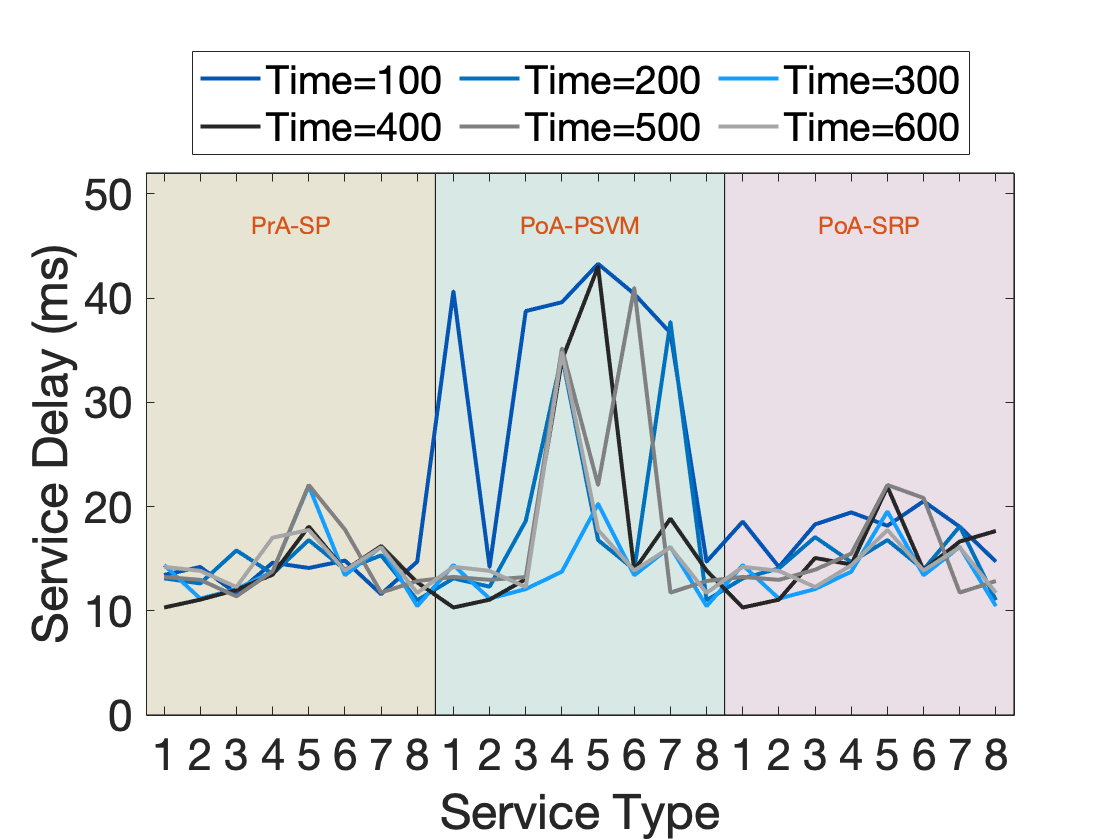}  
		\caption{Rome}
		\label{fig:delay1RM}
	\end{subfigure}
	\begin{subfigure}{.21\textwidth}
		\centering
		\includegraphics[width=1.7in,height=1.2in]{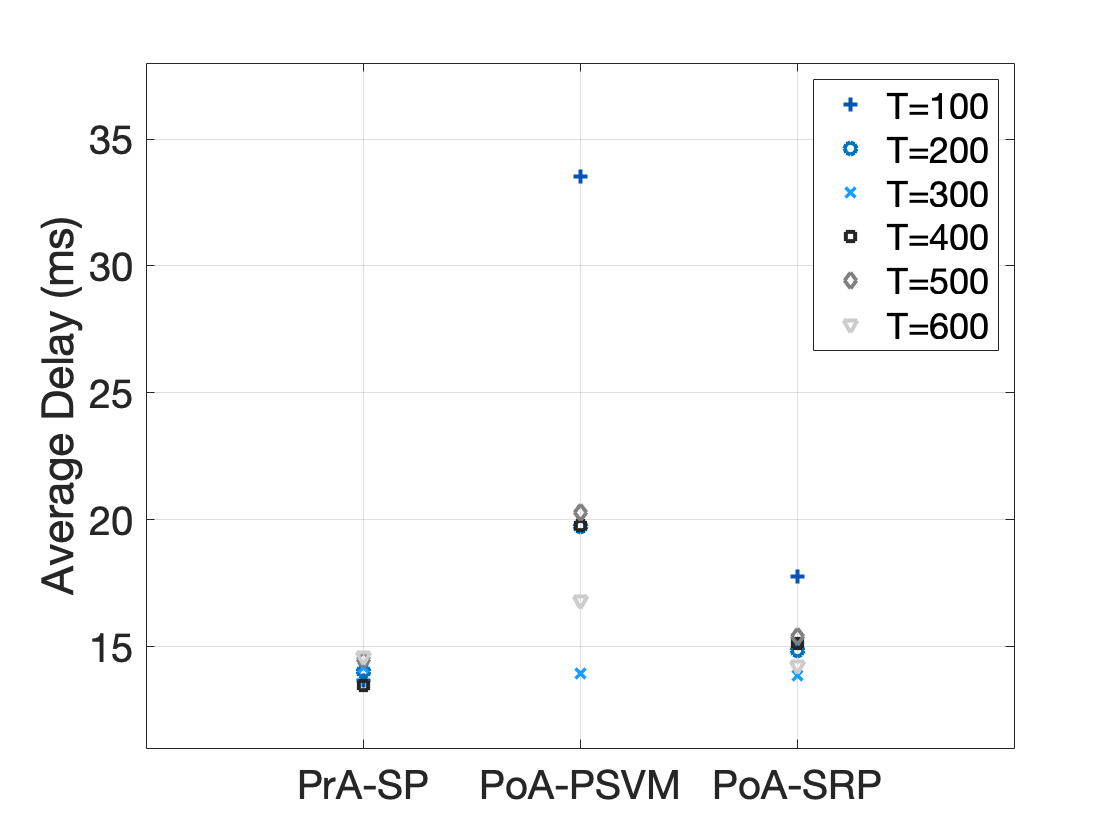}  
		\caption{Rome}
		\label{fig:delay2RM}
	\end{subfigure} 
	\caption{Delay Performance}
	\label{fig:delayBJRM}
\end{figure}
We first compare the delay performance of our framework for different datasets in Fig. \ref{fig:delayBJRM}. Fig. \ref{fig:delay1BJ} and Fig. \ref{fig:delay1RM} depict the service delay for the city environment of Beijing and Rome, respectively. We note that despite the different densities of different vehicular environments, the performance of our proposed framework is consistent and optimal. The service availability is resilient and prompt in PoA-SRP compared to PrA-SP. Further, Fig. \ref{fig:delay2BJ} and Fig. \ref{fig:delay2RM} plots the average delay for the city of Beijing and Rome, respectively. We observe that for Beijing, 50\% of the time, the average delay for PoA-PSVM is fairly similar to PrA-SP and PoA-SRP. On the other hand, for Rome, except for T=100, the average delay is satisfactory and service access is reasonably faster in PoA-PSVM. \par
\begin{figure*}[hbt!]
	\captionsetup[subfigure]{justification=centering}
	\centering
	\begin{subfigure}{.15\textwidth}
		\centering
		\includegraphics[width=1.2in,height=1in]{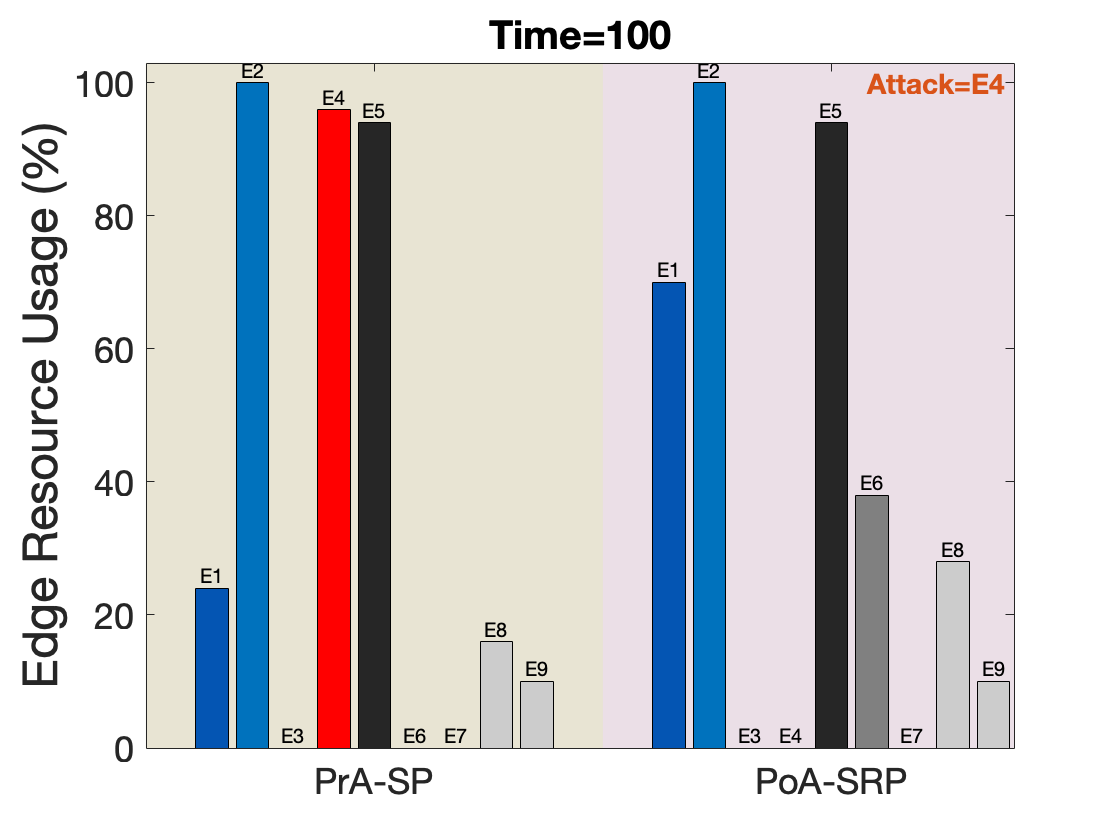}  
		\label{fig:ERU1-BJ}
	\end{subfigure}
	\begin{subfigure}{.15\textwidth}
		\centering
		\includegraphics[width=1.2in,height=1in]{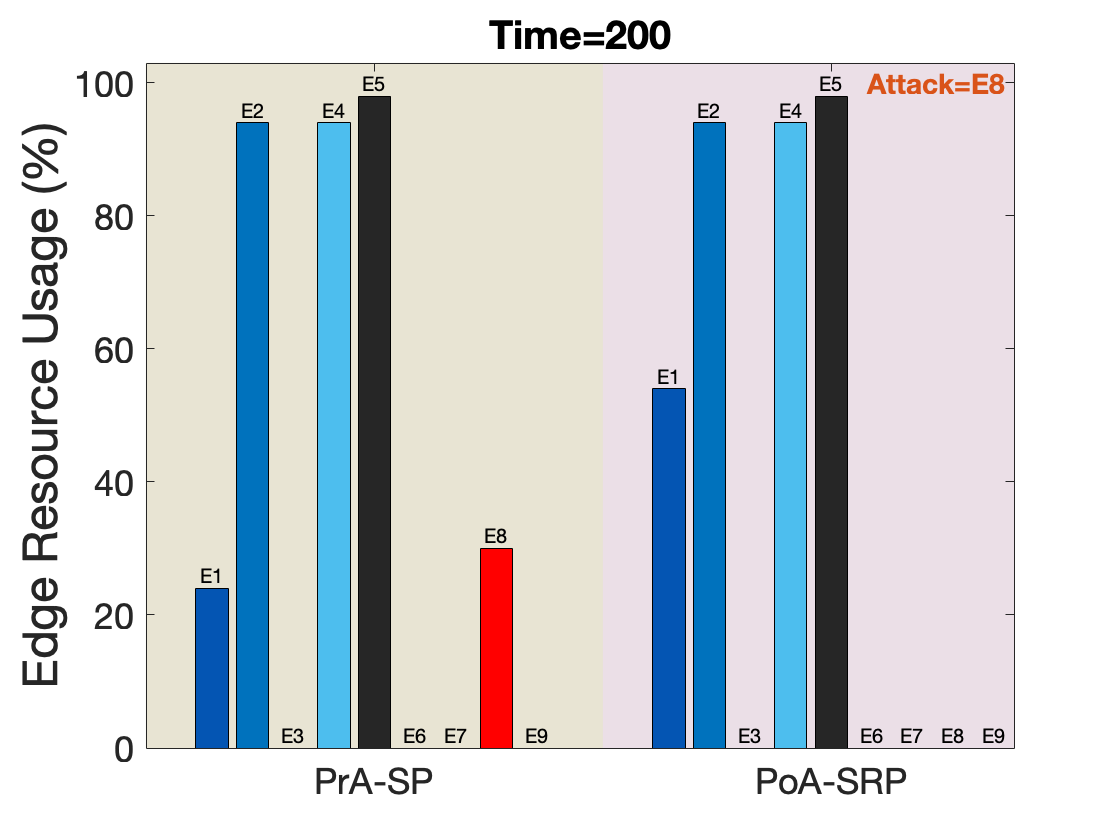}  
		\label{fig:ERU2-BJ}
	\end{subfigure} 
	\begin{subfigure}{.15\textwidth}
		\centering
		\includegraphics[width=1.2in,height=1in]{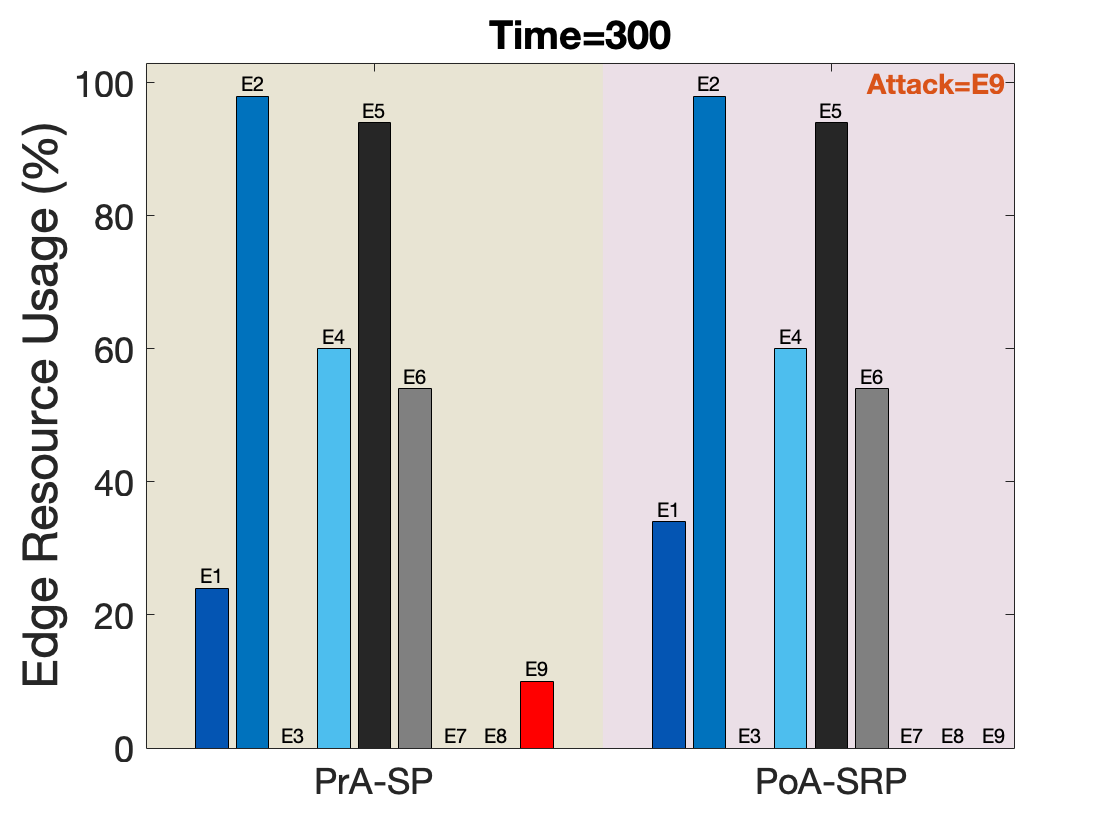}  
		\label{fig:ERU3-BJ}
	\end{subfigure} 
	\begin{subfigure}{.15\textwidth}
		\centering
		\includegraphics[width=1.2in,height=1in]{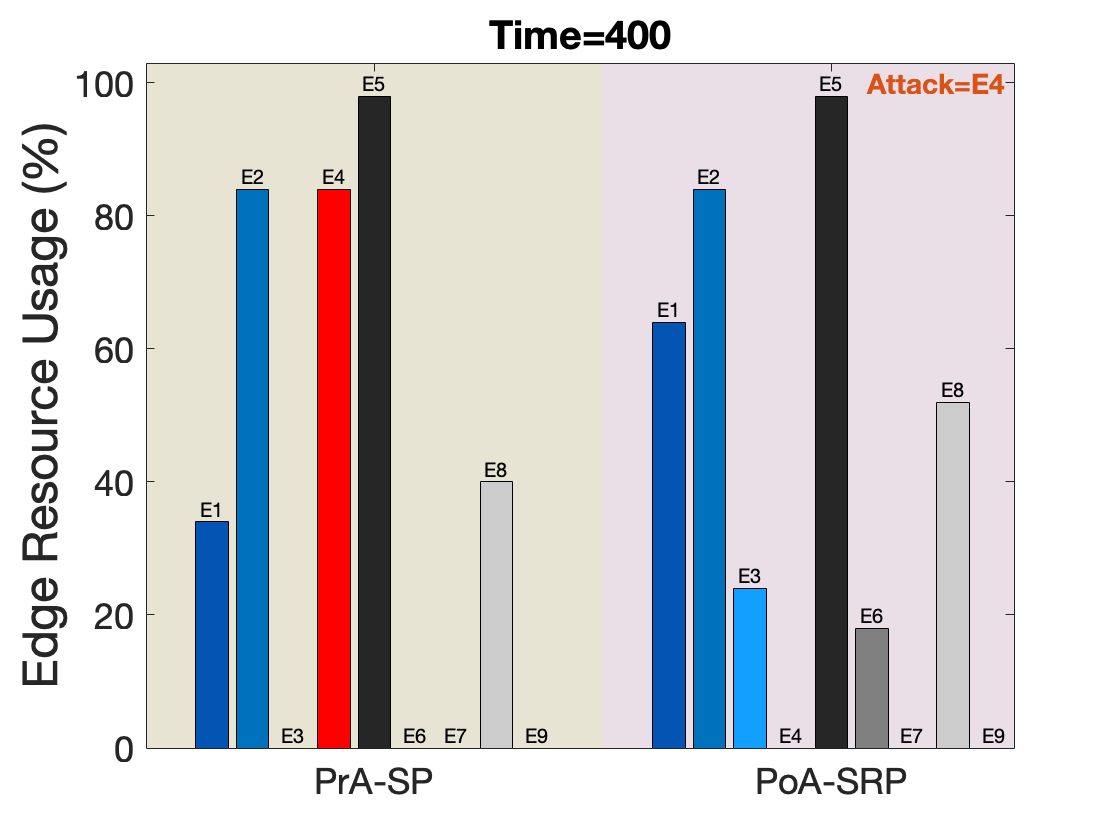}  
		\label{fig:ERU4-BJ}
	\end{subfigure} 
	\begin{subfigure}{.15\textwidth}
		\centering
		\includegraphics[width=1.2in,height=1in]{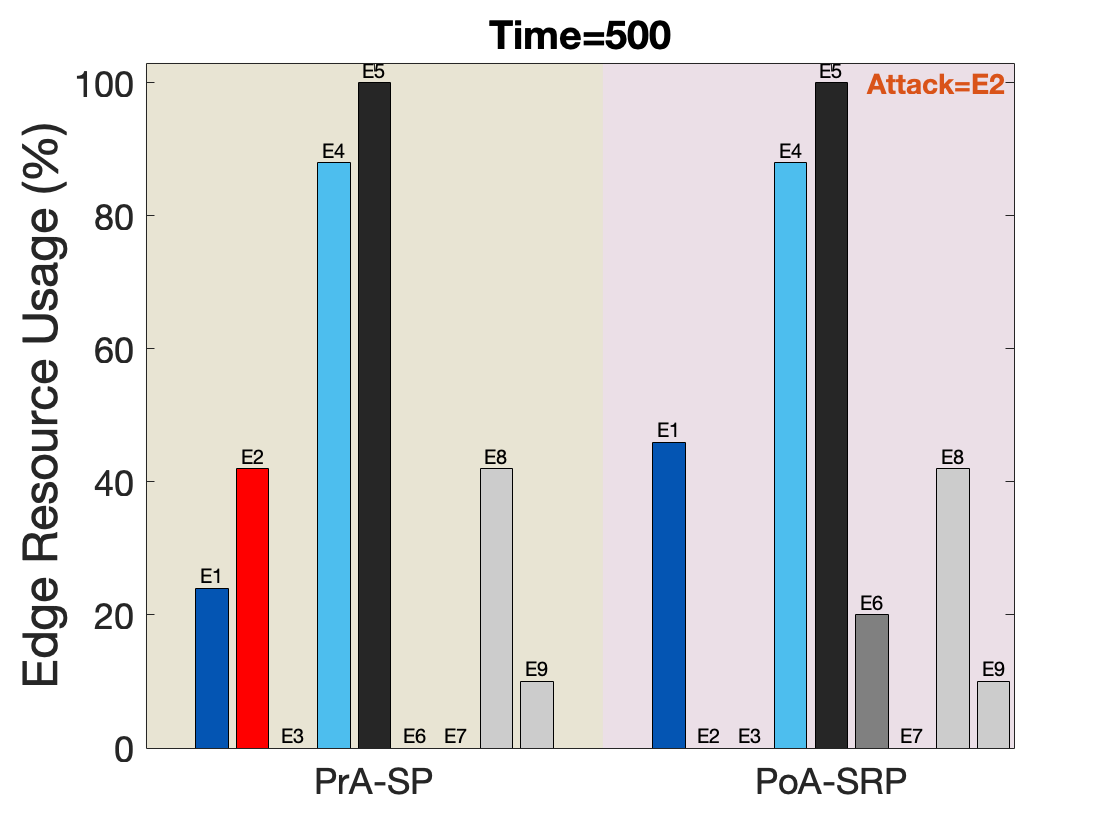}  
		\label{fig:ERU5-BJ}
	\end{subfigure}
	\begin{subfigure}{.15\textwidth}
		\centering
		\includegraphics[width=1.2in,height=1in]{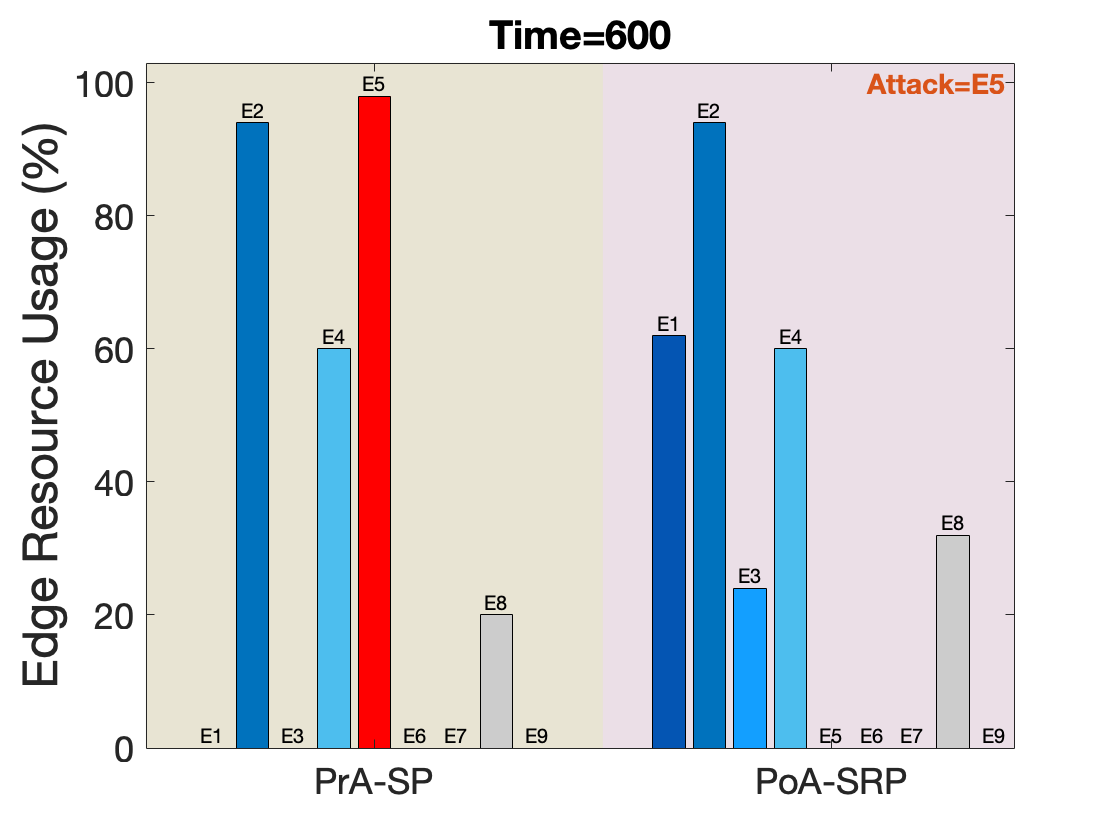}  
		\label{fig:ERU6-BJ}
	\end{subfigure} 
	\caption{Edge resource usage for the Beijing dataset}
	\label{fig:ERU-BJ}
\end{figure*}
\begin{figure*}[hbt!]
	\captionsetup[subfigure]{justification=centering}
	\centering
	\begin{subfigure}{.15\textwidth}
		\centering
		\includegraphics[width=1.2in,height=1in]{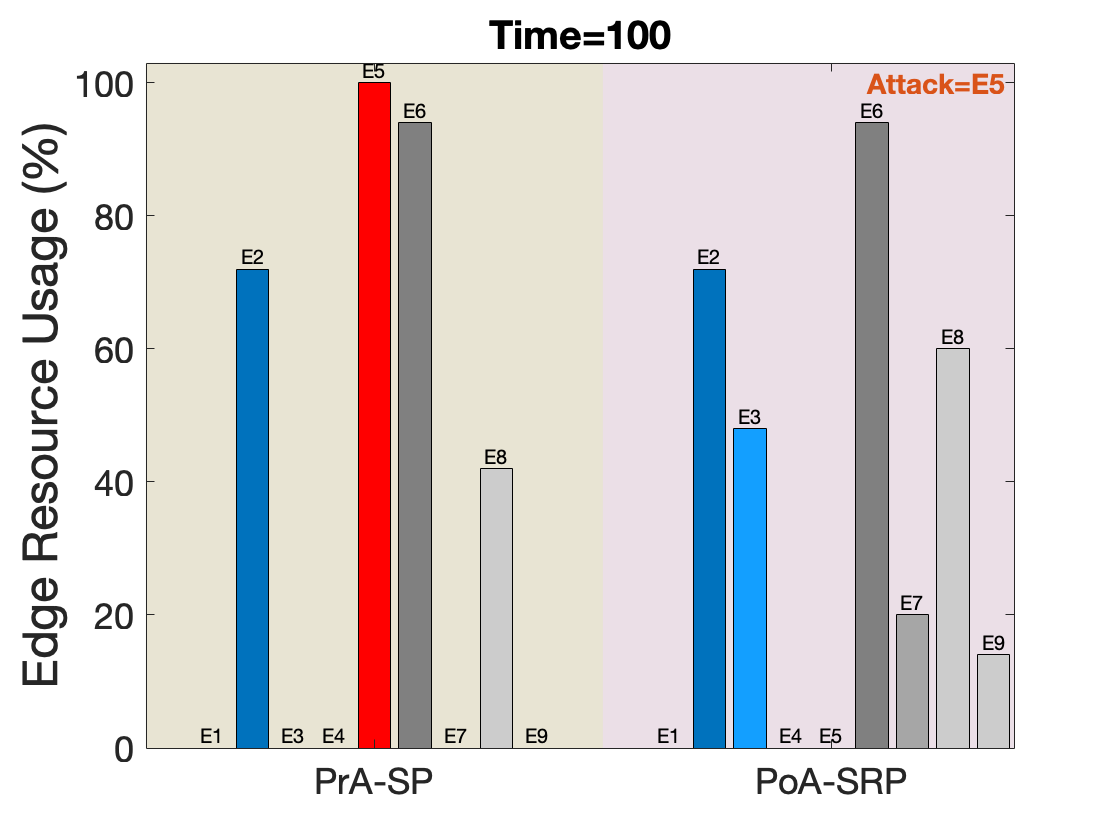}  
		\label{fig:ERU1-RM}
	\end{subfigure}
	\begin{subfigure}{.15\textwidth}
		\centering
		\includegraphics[width=1.2in,height=1in]{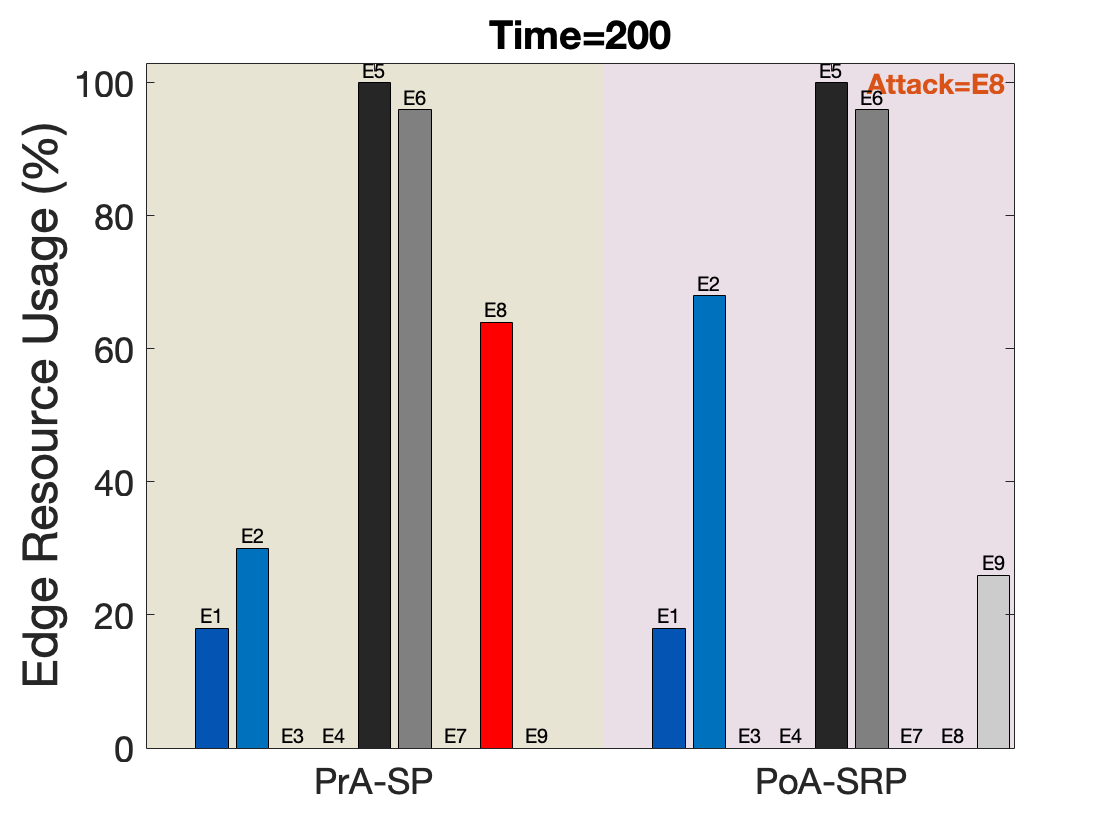}  
		\label{fig:ERU2-RM}
	\end{subfigure} 
	\begin{subfigure}{.15\textwidth}
		\centering
		\includegraphics[width=1.2in,height=1in]{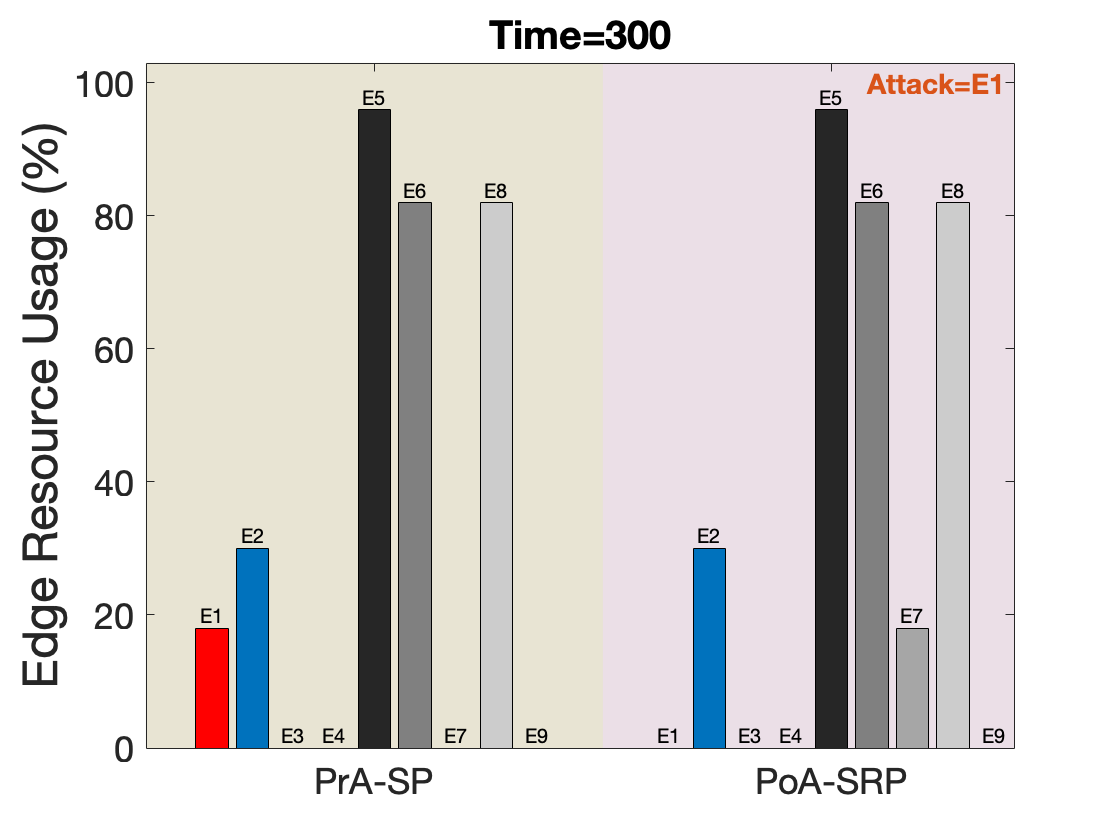}  
		\label{fig:ERU3-RM}
	\end{subfigure} 
	\begin{subfigure}{.15\textwidth}
		\centering
		\includegraphics[width=1.2in,height=1in]{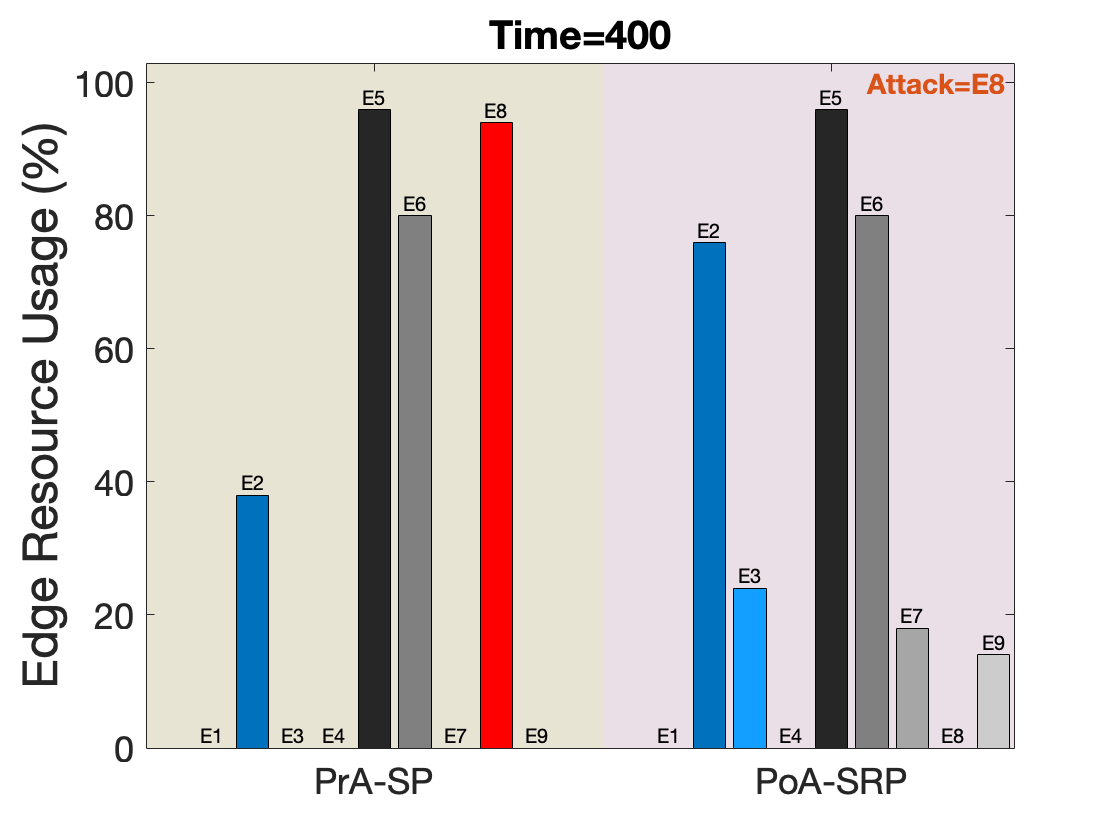}  
		\label{fig:ERU4-RM}
	\end{subfigure} 
	\begin{subfigure}{.15\textwidth}
		\centering
		\includegraphics[width=1.2in,height=1in]{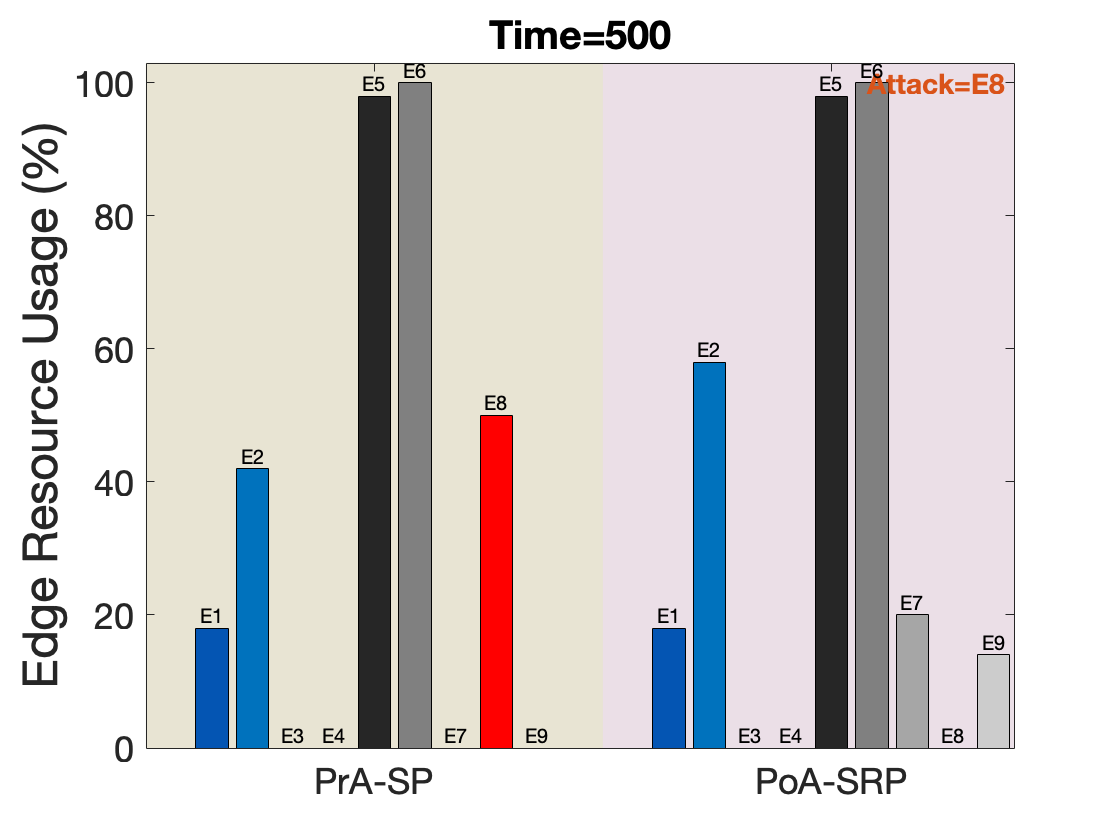}  
		\label{fig:ERU5-RM}
	\end{subfigure}
	\begin{subfigure}{.15\textwidth}
		\centering
		\includegraphics[width=1.2in,height=1in]{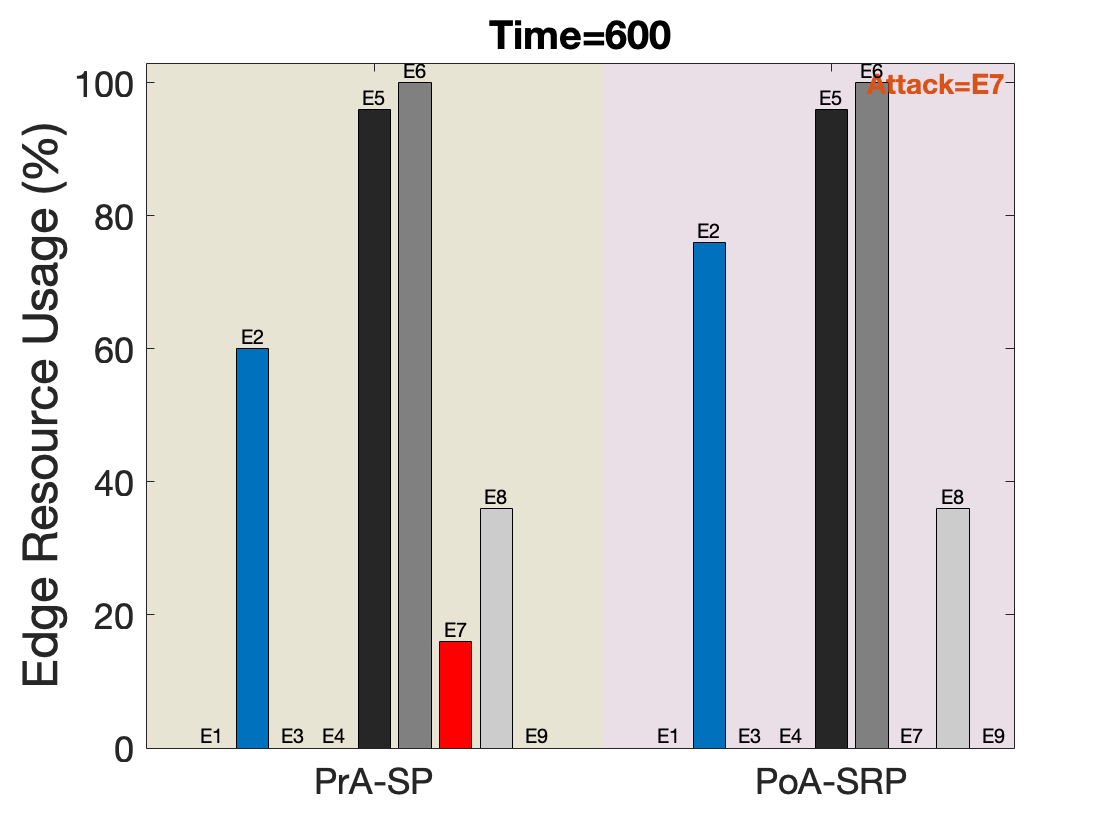}  
		\label{fig:ERU6-RM}
	\end{subfigure} 
	\caption{Edge resource usage for the Rome dataset}
	\label{fig:ERU-RM}
\end{figure*}

We also compare the edge resource usage and the number of active edge nodes for the Rome and Beijing city environment in Fig. \ref{fig:ERU-BJ}-\ref{fig:ERU-RM} and Table \ref{tab:ActiveServers-BJRM}. With the same design parameters, our proposed framework exhibits similar edge resource usage performance to support the resilient attack protection for differently-loaded nodes under attack and with minimal usage of edge resources for different nature of city environment. In Table \ref{tab:ActiveServers-BJRM}, the active number of nodes in pre-attack and post-attack placements are compared for the Beijing and Rome city. We can observe that for the two different vehicular environments the total number of active edge nodes to preserve the service availability is noteworthy and is quite similar in both PrA-SP and PoA-SRP. \par

\begin{table}[htbp]
	\centering
	\caption{Number of active edge nodes}
	\scriptsize
	\tabcolsep=0.09cm
	\begin{tabular}{clcccccc}
		\toprule
		\multicolumn{2}{c}{} & \textbf{T=100} & \textbf{T=200} & \textbf{T=300} & \textbf{T=400} & \textbf{T=500} & \textbf{T=600} \\
		\midrule
		\multirow{2}[4]{*}{\textbf{Beijing}} & \textbf{PrA-SP} & 6     & 5     & 6     & 5     & 6    & 4 \\
		\cmidrule{2-8}          & \textbf{PoA-SRP} & 6     & 4     & 5     & 6     & 6     & 5 \\
		\midrule
		\multirow{2}[4]{*}{\textbf{Rome}} & \textbf{PrA-SP} & 4     & 5     & 5     & 4     & 5     & 5 \\
		\cmidrule{2-8}          & \textbf{PoA-SRP} & 6     & 5     & 5     & 6     & 6     & 4 \\
		\bottomrule
	\end{tabular}%
	\label{tab:ActiveServers-BJRM}%
\end{table}%

\begin{figure}[hbt!]
	\captionsetup[subfigure]{justification=centering}
	\centering
	\begin{subfigure}{.23\textwidth}
		\centering
		\includegraphics[width=1.8in,height=1.25in]{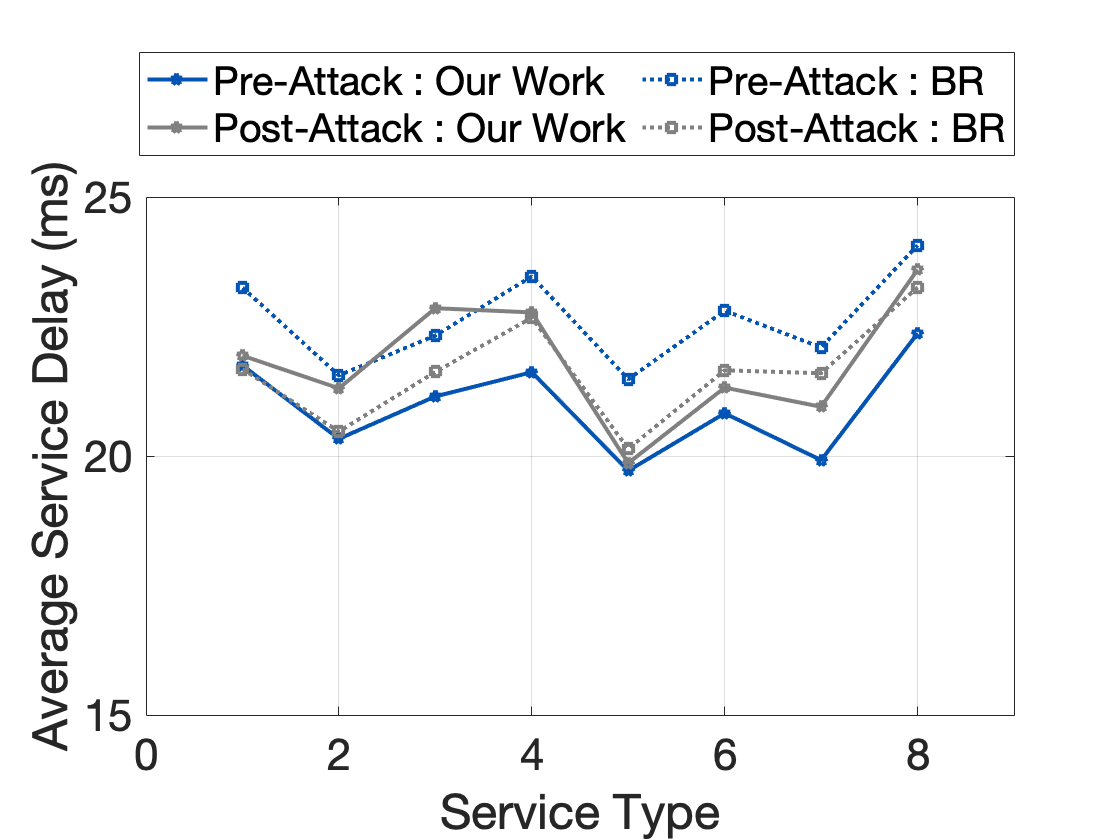}  
		\caption{Beijing}
		\label{fig:CompAvgBJ}
	\end{subfigure}
	\begin{subfigure}{.23\textwidth}
		\centering
		\includegraphics[width=1.8in,height=1.25in]{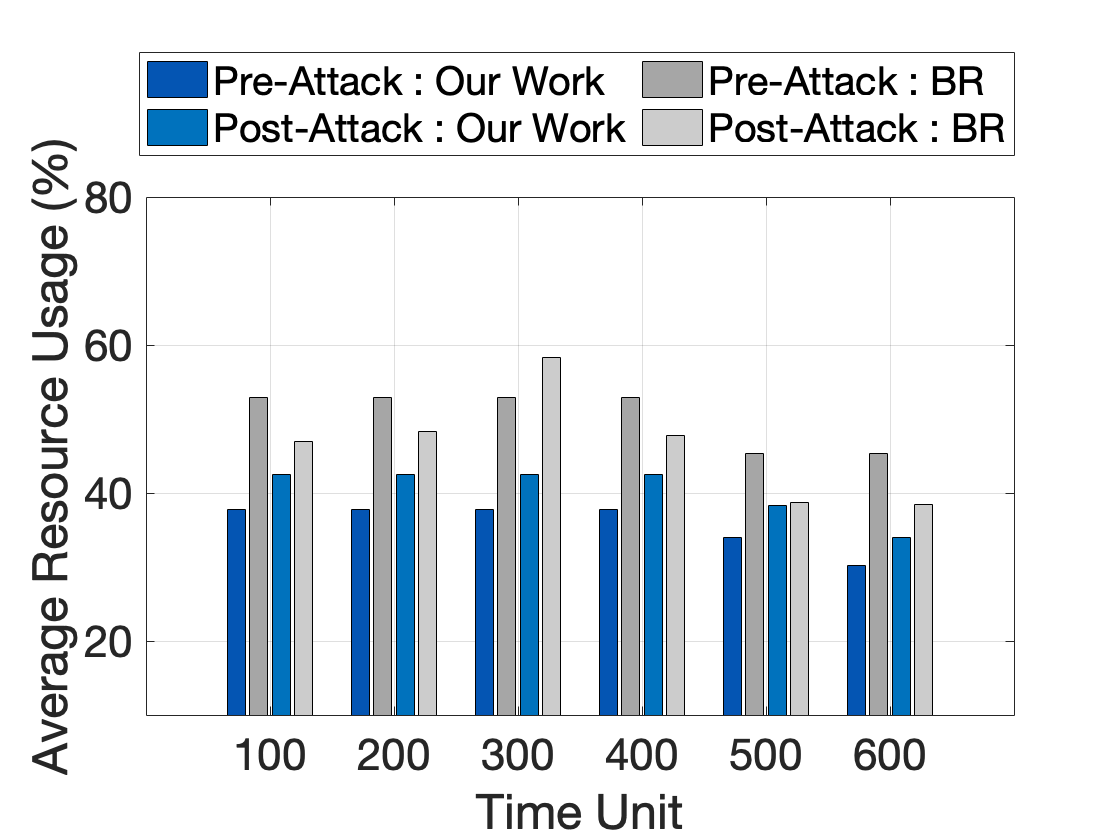}  
		\caption{Beijing}
		\label{fig:CompERUBJ}
	\end{subfigure} \\
	\begin{subfigure}{.23\textwidth}
		\centering
		\includegraphics[width=1.8in,height=1.25in]{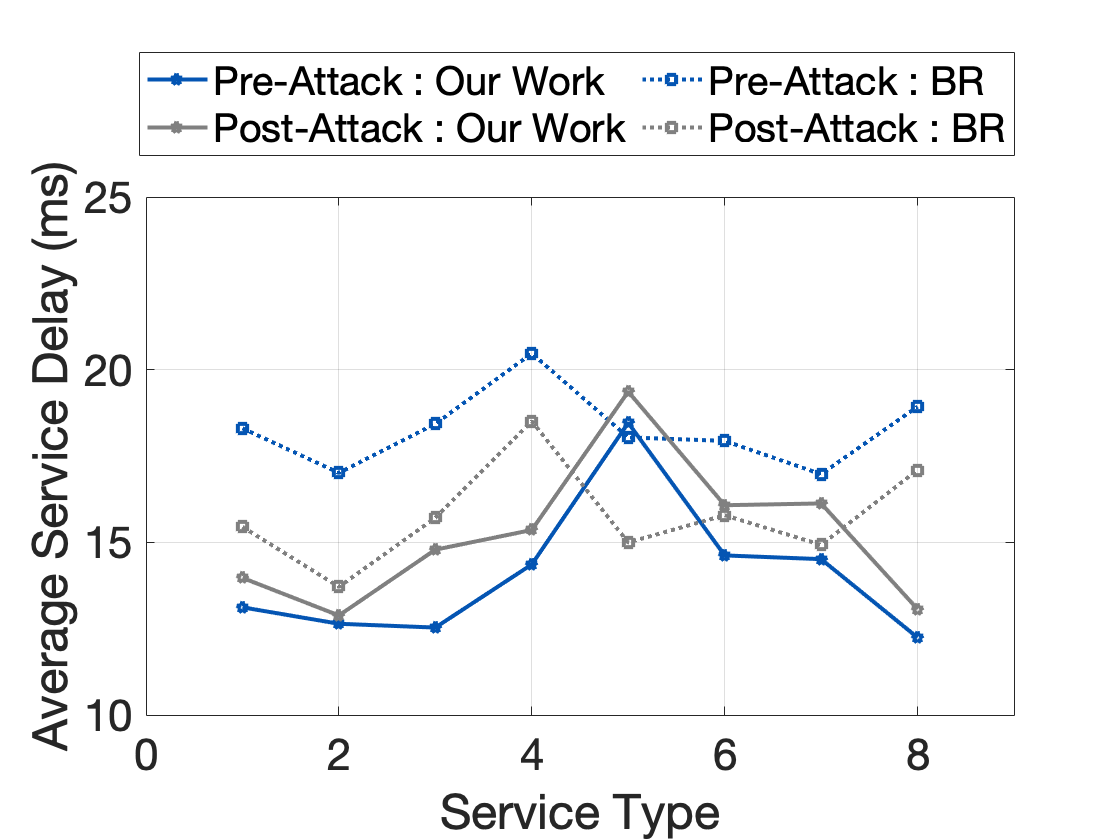}  
		\caption{Rome}
		\label{fig:CompAvgRM}
	\end{subfigure}
	\begin{subfigure}{.23\textwidth}
		\centering
		\includegraphics[width=1.8in,height=1.25in]{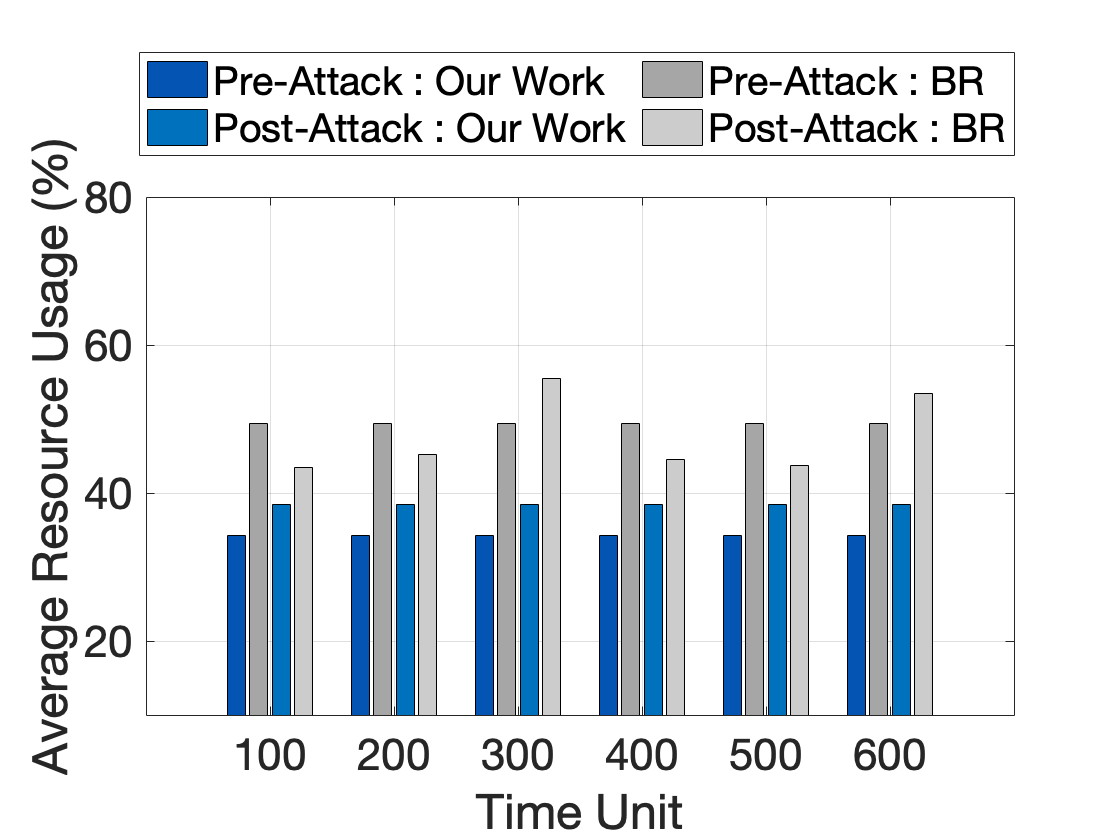}  
		\caption{Rome}
		\label{fig:CompERURM}
	\end{subfigure} 
	\caption{Comparison of our framework with baseline for different datasets}
	\label{fig:delayBJRMComp}
\end{figure}

Next, we evaluate the performance with different datasets against the baseline algorithm in Fig. \ref{fig:delayBJRMComp}. In Fig, \ref{fig:CompAvgBJ} and Fig. \ref{fig:CompAvgRM}, we plot the average service delay experienced by the vehicles in the environment of Beijing and Rome, respectively, to avail different types of services. In general, the average delay for the pre-attack scenario tends to remain higher in BR compared to our proposed resilient service placement framework for both cities with different traffic volumes. In contrast, in the post-attack scenario, the average delay comparison is quite fluctuating. Nevertheless, a fairly similar range of delay values are observed in BR against our work but at the expense of higher edge resources usage. \par

The higher usage of edge resources in BR is verified in Fig. \ref{fig:CompERUBJ} and Fig. \ref{fig:CompERURM} for the city environment of Beijing and Rome, respectively. Despite the changing vehicular environment, the BR method directly impacts the amount of average resources used to ensure the resilient service availability. It can be observed that the average resource usage is always higher in BR with up to 35\%-40\% more usage compared to our method. \par
\begin{table}[htbp]
	\centering
	\scriptsize
	\caption{Run time for different datasets}
	\begin{tabular}{p{2.5em}p{2.3em}p{2.3em}p{2.3em}p{2.3em}p{2.3em}p{2.3em}p{2.3em}p{2.3em}}
		\cmidrule{2-9}    \multicolumn{1}{c}{\multirow{3}[6]{*}{}} & \multicolumn{4}{c}{\textbf{Beijing}} & \multicolumn{4}{c}{\textbf{Rome}} \\
		\cmidrule{2-9}    \multicolumn{1}{c}{} & \multicolumn{1}{c}{\multirow{2}[4]{*}{\textbf{BR}}} & \multicolumn{3}{c}{\textbf{Our Method}} & \multicolumn{1}{c}{\multirow{2}[4]{*}{\textbf{BR}}} & \multicolumn{3}{c}{\textbf{Our Method}} \\
		\cmidrule{3-5}\cmidrule{7-9}    \multicolumn{1}{c}{} &       & \multicolumn{1}{c}{\textbf{PrA-}} & \multicolumn{1}{c}{\textbf{PoA-}} & \multicolumn{1}{c}{\textbf{PoA-}} &       & \multicolumn{1}{c}{\textbf{PrA-}} & \multicolumn{1}{c}{\textbf{PoA-}} & \multicolumn{1}{c}{\textbf{PoA-}} \\
		\multicolumn{1}{c}{} &       & \multicolumn{1}{c}{\textbf{SP}} & \multicolumn{1}{c}{\textbf{PSVM}} & \multicolumn{1}{c}{\textbf{SRP}} &       & \multicolumn{1}{c}{\textbf{SP}} & \multicolumn{1}{c}{\textbf{PSVM}} & \multicolumn{1}{c}{\textbf{SRP}} \\
		\midrule
		\textbf{T=100} & 0.1299 & 0.288 & 0.0422 & 0.0543 & 0.5725 & 0.2762 & 0.0202 & 0.0254 \\
		\midrule
		\textbf{T=200} & 0.1957 & 0.1805 & 0.0243 & 0.0344 & 0.3989 & 0.2947 & 0.0224 & 0.0279 \\
		\midrule
		\textbf{T=300} & 0.1802 & 0.1689 & 0.0213 & 0.0394 & 0.5401 & 0.3139 & 0.0234 & 0.0167 \\
		\midrule
		\textbf{T=400} & 0.1616 & 0.1713 & 0.0196 & 0.0226 & 0.4443 & 0.3139 & 0.0175 & 0.0299 \\
		\midrule
		\textbf{T=500} & 0.1824 & 0.1713 & 0.0242 & 0.0308 & 0.4517 & 0.2800 & 0.0228 & 0.0295 \\
		\midrule
		\textbf{T=600} & 0.1816 & 0.1713 & 0.0235 & 0.0156 & 0.3113 & 0.3157 & 0.0210 & 0.0214 \\
		\bottomrule
	\end{tabular}%
	\label{tab:RumTime-BJRM}%
\end{table}%
Finally, in Table \ref{tab:RumTime-BJRM}, we depict the run time of our method and BR for the dataset of the Beijing and Rome city with varying traffic volumes and mobilities. We note that our proposed framework shows good performance in terms of computation time. The results are consistent with the different datasets and efficient even with the cumulative time of the three ILP models. This is because of a smaller search space in our post-attack ILP models which require smaller run time.

\section{Conclusion}
\label{Sec:Conclusion}
In this paper, we proposed a framework for attack resilient service placement and service availability in  edge-enabled IoV networks and developed three ILP models to jointly solve the problem with a DRL model. The proposed method determines the optimal service placement in the first stage of attack-free scenario taking into account the service delay observed by vehicles and edge resources usage. It also calculates the optimal secondary V2E mappings proactively to maintain disruption-free (or low disruption) service availability to the attack-affected vehicles until recovery takes place. Our model achieves resilience without pre-reserving expensive resources. Upon a failure, we find optimal service recovery placements with the objective of minimizing the edge resource usage. Our proposed solution not only improves the user experience by maintaining low service latency but also reduces the system cost in terms of resource usage and number of active edge nodes. The effect of dynamic traffic changes on service placement and system performance quality is addressed by integrating ILP models with a DRL framework. We carried out extensive performance study to verify the effectiveness of the proposed framework using different datasets representing different city environments.

\balance

% references section
\bibliographystyle{IEEEtran}
\bibliography{IEEEabrv,References}

\vfill

\end{document}